\documentclass[aps,reprint,floatfix, longbibliography]{revtex4-2}

\usepackage{amsmath, braket, amsfonts}

\usepackage[colorlinks,citecolor=black,linkcolor=black,urlcolor=black]{hyperref}

\usepackage{graphicx}

\usepackage[normalem]{ulem}

\begin{document}

\title{Qubit-efficient quantum combinatorial optimization solver}

\author{Bhuvanesh Sundar}
\email[]{bsundar@rigetti.com}
\affiliation{Rigetti Computing, 775 Heinz Avenue, Berkeley, California 94710, USA}

\author{Maxime Dupont}
\affiliation{Rigetti Computing, 775 Heinz Avenue, Berkeley, California 94710, USA}

\begin{abstract}
Quantum optimization solvers typically rely on one-variable-to-one-qubit mapping. However, the low qubit count on current quantum computers is a major obstacle in competing against classical methods. Here, we develop a qubit-efficient algorithm that overcomes this limitation by mapping a candidate bit string solution to an entangled wave function of fewer qubits. We propose a variational quantum circuit generalizing the quantum approximate optimization ansatz. Extremizing the ansatz for Sherrington-Kirkpatrick spin-glass problems, we show valuable properties such as the concentration of ansatz parameters and derive performance guarantees. This approach could benefit near-term intermediate-scale and future fault-tolerant small-scale quantum devices.
\end{abstract}
\maketitle

\section{Introduction}
An ongoing effort in quantum computing seeks to practically demonstrate quantum utility for solving combinatorial optimization problems. Achieving quantum utility could have a significant impact in areas that heavily utilize optimization, such as logistics and supply chains. To this end, various quantum algorithms have been developed and implemented, including annealing~\cite{farhi2000quantum, king2023quantum}, parametrized quantum algorithms with variational optimization~\cite{farhi2014quantum, cerezo2021variational, blekos2024review, zhu2022adaptive}, recursive techniques~\cite{bravyi2020obstacles, brady2024iterative, dupont2023quantum, finvzgar2024quantum}, and others inspired by leading classical approaches~\cite{dupont2024extending, egger2021warm}.

Quantum algorithms typically map each of the problem variables onto individual qubits and recast the original optimization problem as the ground-state search for a Hamiltonian.
One of the obstacles in demonstrating quantum utility is the size of the problems one can address, given that current gate-based quantum computers typically have tens to hundreds of qubits~\cite{harrigan2021quantum, ebadi2022quantum, maciejewski2024design, shaydulin2024evidence, sachdeva2024quantum, moses2023race, pelofske2024short, dupont2023quantum, montanez2025toward}, and quantum annealers have thousands of qubits~\cite{king2023quantum}~\footnote{However, the effective problem sizes solved on quantum annealers are smaller than the number of physical qubits due to embedding overheads~\cite{choi2008minor}.}. Indeed, real-world problems typically have several thousand variables, which can be handled by classical solvers better than quantum solvers today.
A prevailing practical challenge in adding more qubits to the computers is hardware noise, which limits their solution quality. In the absence of quantum error correction~\cite{roffe2019quantum}, more qubits generally means larger and therefore noisier quantum programs, compounding toward a deteriorating output. Meanwhile, initial error-corrected quantum computers are expected to have fewer logical qubits than necessary for quantum utility in optimization. Therefore, qubit-efficient methods for optimization problems could benefit near-term intermediate-scale as well as fault-tolerant small-scale quantum computers~\cite{bluvstein2024logical, google2023suppressing, google2025quantum, da2024demonstration, krinner2022realizing}.

Several qubit-efficient methods have been proposed to overcome the size of current quantum computers. Some of these are generic methods to break down a large quantum circuit into an equivalent set of smaller quantum circuits that can be executed on current devices~\cite{bravyi2016trading,peng2020simulating,tang2021cutqc,decross2023qubit,fang2025dynamic,brandhofer2023optimal,hua2023caqr,kim2023evidence}, and have been used to solve large optimization problems with more variables than qubits~\cite{bechtold2023investigating,dupont2025benchmarking}---at the expense of large computational time overheads in classical or quantum resources. Some approaches, specific to combinatorial optimization, employ decompositions that stitch together an approximate solution for a large optimization problem by solving smaller subproblems~\cite{angone2023hybrid, bach2024mlqaoa, liu2022hybrid}, and again have large overheads induced by the decomposition. Other strategies employ nontrivial mappings between classical variables of the optimization problem and qubits on the device, such as amplitude encoding~\cite{ranvcic2023noisy, chatterjee2024solving, tene2026variational}, correlation encoding~\cite{sciorilli2025towards}, and multibasis encodings~\cite{patti2022variational}, each with some challenges for quantum utility. For instance, the amplitude encoding strategy is classically simulatable, while the other strategies use hardware-native variational circuits that may be difficult to optimize~\cite{bittel2021training, larocca2025barren}.

Here, we develop a qubit-efficient encoding based on a many-to-one mapping from variables to qubits. In this approach, classical bit strings are encoded as entangled wave functions of fewer qubits. Additionally, we physically motivate a variational quantum algorithm for this encoding, inspired by the quantum approximate optimization algorithm (QAOA)~\cite{farhi2014quantum}. While the QAOA uses $q=N$ qubits to store $N$ variables~\cite{farhi2014quantum}, the proposed algorithm makes the number of qubits $q$ a tunable parameter. Compared to the overhead of previous decomposition methods~\cite{angone2023hybrid, bach2024mlqaoa, liu2022hybrid, bravyi2016trading,peng2020simulating,tang2021cutqc,decross2023qubit,brandhofer2023optimal,kim2023evidence, bechtold2023investigating,dupont2025benchmarking}, this algorithm only introduces a polynomial overhead in $Np$ over the QAOA for $p$ layers.
We show that this algorithm has several properties also present in the QAOA, such as parameter clustering and performance guarantees for some classes of problems~\cite{farhi2022quantum, wurtz2021fixed, wurtz2021maxcut, basso2022quantum}. We numerically benchmark the algorithm for Sherrington-Kirkpatrick (SK) spin-glass models~\cite{sherrington1975solvable} and test it on Rigetti's Ankaa\textsuperscript{TM}-9Q-3 superconducting chip. Finally, we numerically investigate an adaptive version of the variational ansatz, where the ansatz is dynamically constructed, analogous to ADAPT-QAOA~\cite{zhu2022adaptive}.

\begin{figure}[!t]\centering
\includegraphics[width=0.9\columnwidth]{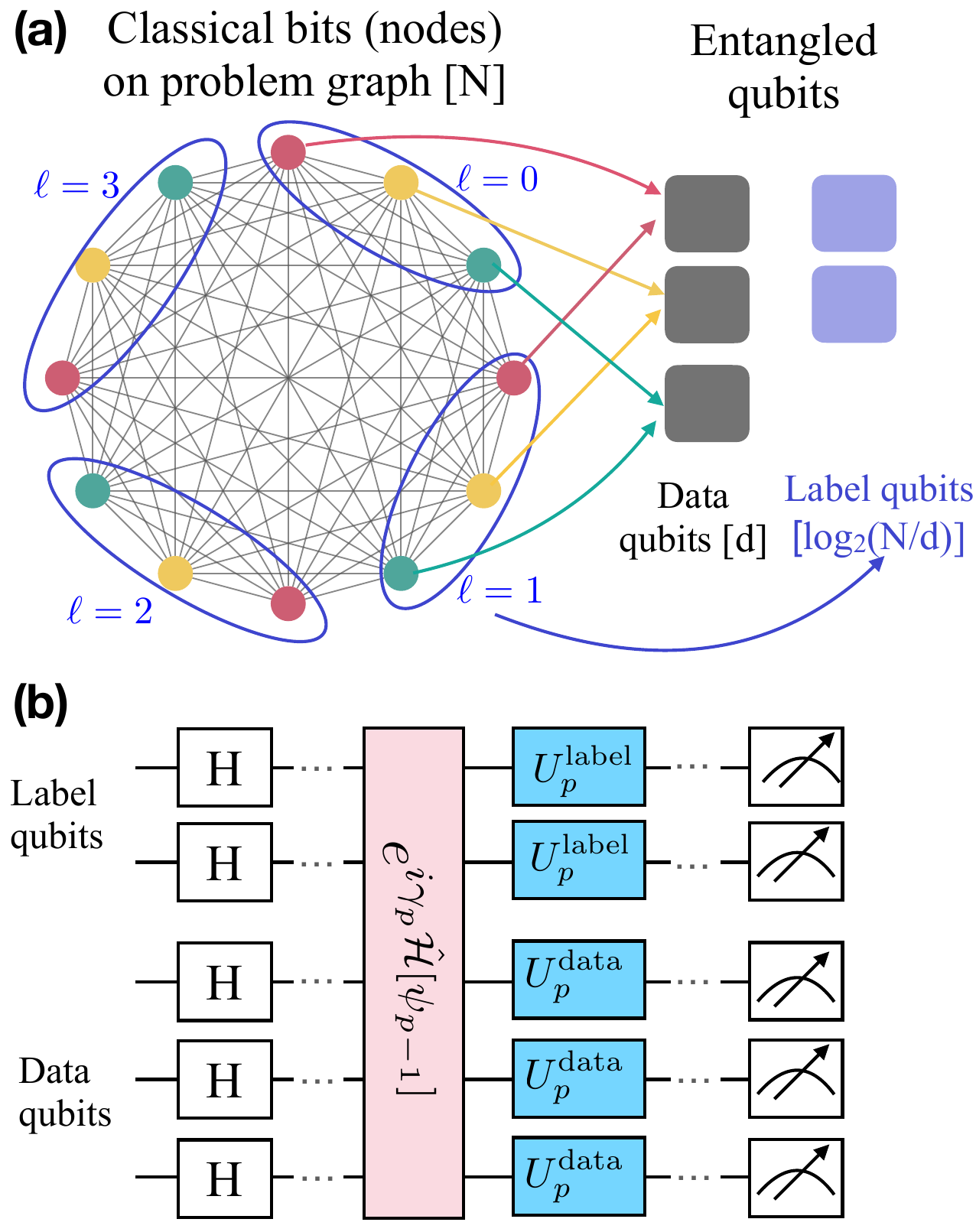}
\caption{(a) Qubit-efficient encoding of variables onto qubits via a many-to-one mapping from variables to qubits. Graph nodes, which host variables, are color-coded for convenience. All nodes with the same color are mapped to the same \textit{data} qubit (in gray). The value of the \textit{label} qubits (purple) dictates which node is stored in the data qubit. Links between nodes represent interactions between nodes in the optimization problem [see text]. (b) parametrized quantum circuit to solve combinatorial optimization problems with qubit-efficient encoding. The circuit consists of an initial layer of Hadamard gates, followed by alternating layers of entangling ($e^{-i\gamma_p\hat{\mathcal{H}}}$) and one-qubit gates ($U_p$). The entangling gates in the $p$th layer are determined from measurements of the wave function at the end of the $(p-1)$th layer. The parametric angles are chosen classically to minimize the cost measured by the quantum computer.
}
\label{fig:encoding+ansatz}
\end{figure}

\section{Qubit-efficient encoding}\label{sec: encoding} Given $N$ binary variables $\boldsymbol{z} = (z_1, \cdots, z_N)$, we divide the variables into $N/d$ groups of $d$ variables each, as shown in Fig.~\ref{fig:encoding+ansatz}(a), and denote each group with a label $\ell$ that runs from 0 to $N/d-1$. $d$ can be chosen by the user. Without loss of generality, we assume $N$ is divisible by $d$. 

The $N$ variables in $\boldsymbol{z}$ can be grouped in any way into $(N/d)$ groups $\boldsymbol{z}_\ell$ with $d$ variables each. Thus, each variable $i\in [1,N]$ is identified by a label $\ell_i$ that labels its group, and a data index $d_i$ that indicates its location in the group $\mathbf{z}_{\ell_i}$. For concreteness, we consider the grouping with the first $d$ variables in $\boldsymbol{z}_1 = (z_1, z_2, \cdots, z_d)$, the next $d$ variables in $\boldsymbol{z}_2 = (z_{d+1} \cdots z_{2d})$, and so on. For this choice of grouping, we have $\ell_i = \lfloor \frac{i}{d} \rfloor$ and $d_i = i\ \textrm{mod}\ d$.

We store the $N$ variables in an entangled wave function $\ket{\phi(\boldsymbol{z})}$ of $q = d + \log_2(N/d)$ qubits as
\begin{equation}\label{eq:dense-encoding}
 \ket{\phi(\boldsymbol{z})} \equiv \sum\nolimits_{\ell=0}^{N/d-1} \lambda_\ell \ket{\ell}_{\rm label}\otimes\ket{\boldsymbol{z}_\ell}_{\rm data},
\end{equation}
where $\ket{\ell}$ is the $\log_2(N/d)$-qubit computational basis state representing the bit string for $\ell$, $\ket{\boldsymbol{z}_\ell} $ stores the $d$ binary variables in $\mathbf{z}_\ell$ defined above, and $\lambda_\ell$ are arbitrary complex coefficients with $\sum_\ell |\lambda_\ell|^2=1$. This wave function stores all $N$ binary variables of $\boldsymbol{z}$. A single projective measurement only gives the values of $d$ variables. For example, in a projective measurement, if the \textit{label} qubits were measured as the bit-string representation of $\ell$, then the variables in $\boldsymbol{z}_\ell \equiv (z_{(N/d-1)\ell+1}, z_{(N/d-1)\ell+2}, \cdots, z_{(N/d)\ell})$ are assigned the values measured in the \textit{data} qubits.

In the case of $d=N$, $\ket{\phi(\boldsymbol{z})}\equiv\ket{z_1, z_2, \cdots, z_N}_{\textrm data}$ reduces to the usual way of storing binary variables in a product state of the qubits. This is the most common approach for quantum combinatorial optimization~\cite{cerezo2021variational, blekos2024review}. For $d<N$, the wave function stores the variables using a $(N/d)$-to-one map between the variables and data qubits. This encoding describes a generic method to efficiently store classical binary variables in qubits.

Figure~\ref{fig:encoding+ansatz}(a) shows an example of the qubit-efficient encoding with $N=12, d=3$. In this case, the variables $\boldsymbol{z} = (z_1, \cdots, z_{12})$ are grouped into four groups as $\boldsymbol{z_0} = (z_1, z_2, z_3)$, $\boldsymbol{z_1} = (z_4, z_5, z_6)$, $\boldsymbol{z_2} = (z_7, z_8, z_9)$, and $\boldsymbol{z_3} = (z_{10}, z_{11}, z_{12})$.
Variables within one group, $\boldsymbol{z_\ell}$, are circled by a blue ellipse in Fig.~\ref{fig:encoding+ansatz}(a), and are given the label $\ell$.
All variables $z_i$ with the same data qubit $d_i$ are given the same color, for convenience of visualization. In this case, the 12 variables are stored in five qubits. The qubit-efficient wave function in this case is
\begin{align}
\ket{\phi(\mathbf{z})} &= \lambda_0 \ket{00}_{\rm label}\ket{z_1z_2z_3}_{\rm data} + \lambda_1 \ket{01}_{\rm label}\ket{z_4z_5z_6}_{\rm data}\nonumber\\
& + \lambda_2 \ket{10}_{\rm label}\ket{z_7z_8z_9}_{\rm data} + \lambda_3 \ket{11}_{\rm label}\ket{z_{10}z_{11}z_{12}}_{\rm data}.
\end{align}

\section{Connections to previous qubit-efficient methods}

\begin{table*}
\centering
\begin{tabular}{|c|c|c|c|c|c|}
    
    \hline
    Problem type & \# variables & \# qubits & Circuit depth & Approximation ratio & Source\\
    \hline
    Random QUBO & 64 & 7 & 4 & 0.73 & Fig.3 of \cite{tan2021qubit}\\
    Random QUBO & 64 & 7 & 20 & 0.88 & Fig.3 of \cite{tan2021qubit}\\
    Random QUBO & 42 & 7 & 6 & 0.778 & Fig.7 of \cite{tan2021qubit}\\
    3reg. Maxcut & 128 & 7 & 1 & 0.71 -- 0.81 & Table 2 of \cite{ranvcic2023noisy}\\
    Maxcut (graph density=0.3) & 64 & 6 & 1 & 0.898 & Table 1 of \cite{chatterjee2024solving}\\
    9reg. Maxcut & 64 & 6 & 50 & 1 & Table 1 of \cite{tene2026variational}\\
    Weighted 3reg. Maxcut & 512 & 13 & 16 & 0.96 & Fig.~6 of \cite{podobrii2026qubit}\\
    Sherrington-Kirkpatrick & 64 & 7 & 4 & 0.77 & Fig.~\ref{fig:result} of this work\\
    \hline
\end{tabular}
\caption{Comparison of approximation ratios~\footnote{The approximation ratio in Table~\ref{table} is defined as $r_{\rm table} \equiv (C_{\rm max}-C)/(C_{\rm max}-C_{\rm min})$. The relation of the approximation ratio in Table~\ref{table} and that used elsewhere in this paper is $r_{\rm table} = (1+r)/2$.} produced by previous works that used exponentially efficient methods for solving various optimization problems. Tan \textit{et al.}~\cite{tan2021qubit} report average approximation ratios for three problem instances, whereas the other works solve only one problem instance. The result from this work, reported on the last line of the table, is obtained from an average of five problem instances, and has been rescaled to match the definition of approximation ratio in other works (see footnote).
}
\label{table}
\end{table*}

Before we proceed to describe how to use this qubit-efficient encoding to solve optimization problems, we describe a few other methods in the literature that have used qubit-efficient techniques to solve optimization problems. We tabulate exemplary results of quantum optimization with these methods in Table~\ref{table}.

The case of $d=1$ was first proposed in~\cite{tan2021qubit} and further explored in~\cite{huber2024exponential, leonidas2024qubit, azad2022solving, perelshtein2023nisq}, and is called \textit{minimal encoding} there. This case is notable since it requires only $1+\log_2(N)$ qubits to store $N$ binary variables.
Amplitude encoding is another method that uses $\log_2(N)$ qubits to store $N$ binary variables, and has also been used to solve optimization problems in Refs.~\cite{tene2026variational, ranvcic2023noisy, chatterjee2024solving}.

Despite the number of qubits being exponentially fewer in the amplitude encoding in these works, and in the minimal encoding of Ref.~\cite{tan2021qubit}, the methods are conceptually different. In the minimal encoding ($d=1$) presented in this work and Refs.~\cite{tan2021qubit, leonidas2024qubit, azad2022solving, perelshtein2023nisq}, the variables are stored in one data qubit. Each projective measurement of the wave function samples one variable, and the value of that variable is binary. In amplitude encoding, the variables are stored in the \textit{amplitudes} of the wave function. Projective measurements of the wave function \textit{do not} directly provide the values of the variables. Instead, the cost function is obtained from an expectation value of the Laplacian matrix of the graph. As a corollary, the amplitude encoding method does not appear to be generalizable to cost functions that have higher order than quadratic, whereas we describe how to estimate arbitrary cost functions with the qubit-efficient method in Sec.~\ref{sec: cost}.

The minimal encoding of Ref.~\cite{tan2021qubit} and amplitude encoding have been used to solve maximum cut and quadratic unconstrained binary optimization (QUBO) optimization problems on small ensembles of problems of various sizes. Tan \textit{et al.}~\cite{tan2021qubit} used a hardware-native ansatz to solve QUBO problems with random coefficients, Ref.~\cite{tene2026variational} used a hardware-independent all-to-all ansatz to solve maximum cut, and Refs.~\cite{ranvcic2023noisy, chatterjee2024solving} used a diagonal ansatz.

Tan \textit{et al.}~\cite{tan2021qubit} also introduced a method to go beyond the minimal encoding ($d=1$) case. However, their method to estimate the cost function is different from our method (see Sec.~\ref{sec: cost}), and instead their method is more closely related to the correlation encoding of Ref.~\cite{sciorilli2025towards}, with additional steps to resolve conflicts between multiple values for the same variable.

Podobrii and colleagues~\cite{podobrii2026qubit} proposed to use amplitude encoding to encode groups of classical variables to be flipped by a local-search heuristic. In this method, choosing a reference bit string $\boldsymbol{z}^{(0)}$, they decide which groups of variables to flip, to minimize the cost. Denoting the number of groups that could be flipped as $\ell$, their method needed $\log_2(\ell)$ qubits, where $\ell$ can be tuned from $N$ to $2^N$. For example, a 1-local-search heuristic can choose to flip one out of $N$ variables, and this choice is encoded in $\log_2 N$ qubits, whereas a search heuristic that can flip groups of all sizes $r$, $0 \leq r \leq N$, has $2^N$ groups and therefore requires $N$ qubits.

A sampling of results from all the methods described above is presented in Table~\ref{table}.

We note that quantum circuits with exponentially fewer qubits can be classically simulated in time scaling as $O(Np)$ where $p$ parametrizes the circuit depth (see below), and similarly for $d\lesssim \log_2 N$. Under the widely held belief that quantum algorithms may not be able to achieve exponential speedup for NP-hard problems, it is unclear where the difficulty for the quantum algorithms at $d\lesssim \log_2N$ lies, with potential candidates being a large circuit depth $p$ or expensive computational time for parameter search.

Here, our main aim is to develop a quantum algorithm that uses qubit-efficient encoding to solve combinatorial optimization problems, and investigate the cases with $1\leq d\leq N$.

\section{Expressing a binary optimization problem}\label{sec: cost}
Binary optimization problems seek to minimize a cost function $C(\boldsymbol{z})$ of variables $\boldsymbol{z} = (z_1, \cdots, z_N)$ with $z_i\in\{\pm1\}$, where the function $C$ is generically written as a polynomial in the variables, e.g., $C = \sum_i w_i z_i + \sum_{ij} w_{ij} z_iz_j + \cdots$. To develop a quantum algorithm that minimizes this cost, one should express the cost as expectation value(s) with respect to samples from the wave function $\ket{\psi}$. This task is trivial for $d=N$, but nontrivial for $d<N$.

For simplicity, let us consider one of the singleton terms in the cost, $w_i z_i$. Since the value of the variable $z_i$ can be obtained only when the label bits are measured as $\ell_i$, an estimation $\overline{z}_i$ for this variable is obtained from averaging the shot values of the Pauli operator $\hat Z_{d_i}$, postselecting only those measurements with the label bits as $\ell_i$. This can be written as the conditional expectation value,
\begin{equation}\label{eq:zbar}
 \bar{z}_i = \frac{\braket{\psi \vert \hat P_{\ell_i}\hat Z_{d_i} \vert \psi}}{\braket{\psi \vert \hat P_{\ell_i} \vert \psi}},
\end{equation}
where $\hat P_\ell = \ket{\ell}_{\textrm{label}}\bra{\ell}_{\textrm{label}}$ is a projection operator.
Similarly, for variables $z_i$ and $z_j$ with the same label, $\ell_i = \ell_j$, $z_i z_j$ can be estimated by postselecting on shots where the label bits were $\ell_i$, therefore the estimate $\overline{z_i z_j}$ is
\begin{equation}
 \left(\overline{z_i z_j}\right)_{\ell_i = \ell_j} = \frac{\braket{\psi \vert \hat P_{\ell_i}\hat Z_{d_i}\hat Z_{d_j} \vert \psi}}{\braket{\psi \vert \hat P_{\ell_i} \vert \psi}}.
 \label{eq:zzbar}
\end{equation}
If $\ell_i \neq \ell_j$, $z_i$ and $z_j$ are never measured together in the same shot, therefore we estimate $\overline{z_i z_j}$ as $\bar{z}_i \times \bar{z}_j$.
In this way, quadratic terms of the form $C(\boldsymbol{z}) = \sum_{i<j} w_{ij} z_i z_j$, for example, can be expressed as
\begin{equation}\label{eq:cost}
C[\psi] = \sum_{\substack{i<j\\ \ell_i=\ell_j}} w_{ij} \overline{z_i z_j} + \sum_{\substack{i<j\\ \ell_i \neq \ell_j}} w_{ij} \bar{z}_i \bar{z}_j.
\end{equation}
This construction can be straightforwardly extended to measure any polynomial in $\boldsymbol{z}$. 

If the wave function $\ket{\psi}$ encodes a single bit string $\boldsymbol{z}$, i.e. $\ket{\psi}$ is of the form of $\ket{\phi(\boldsymbol{z})}$ in Eq.~\eqref{eq:dense-encoding}, then it can be shown that the conditional expectation values in Eqs.~\eqref{eq:zbar} and~\eqref{eq:zzbar} are $\overline{z}_i = z_i$ and $(\overline{z_i z_j})_{\ell_i = \ell_j} = z_i z_j$, and therefore $C(\boldsymbol{z}) = C[\ket{\phi(\boldsymbol{z})}]$.

For a general $\ket{\psi}$, the cost in Eq.~\eqref{eq:cost} can be written as $C[\psi] = \braket{\psi \vert \hat{\mathcal{H}}[\psi] \vert \psi}$ with
\begin{align}\label{eq:H}
 \hat{\mathcal{H}}[\psi] = & \sum_{\substack{i<j\\ \ell_i=\ell_j}} w_{ij} \frac{ \hat P_{\ell_i}\hat Z_{d_i}\hat Z_{d_j} }{\braket{\psi \vert \hat P_{\ell_i} \vert \psi}}
\nonumber\\
+ & \frac{1}{2}\sum_{\substack{i<j\\ \ell_i\neq\ell_j}} w_{ij} \left(\bar{z}_i \frac{ \hat P_{\ell_j}\hat Z_{d_j} }{\braket{\psi \vert \hat P_{\ell_j} \vert \psi}} 
+ \bar{z}_j \frac{ \hat P_{\ell_i}\hat Z_{d_i} }{\braket{\psi \vert \hat P_{\ell_i} \vert \psi}}\right)
\end{align}
where $\bar{z}_i$ was defined in Eq.~\eqref{eq:zbar}.
We highlight that the Hamiltonian $\hat{\mathcal{H}}[\psi]$ depends on $\ket{\psi}$, a consequence of the cost being nonlinear in $\ket{\psi}\bra{\psi}$. 

We point out that there is no direct quantum coupling between $z_i$ and $z_j$ with different labels ($\ell_i \neq \ell_j$), in $\hat{\mathcal{H}}$. There is only a mean field coupling between these variables. However, we show that this does not limit the performance of the ansatz, since the cost that is being minimized still includes the relevant terms.

The above construction of $\hat{\mathcal{H}}$ and $C$ smoothly connects in the $d=N$ limit with their standard forms in the QAOA, $\hat{\mathcal{H}} = \sum_i w_i \hat Z_i + \sum_{ij} w_{ij} \hat Z_i\hat Z_j + \cdots$ and $C = \braket{\hat{\mathcal{H}}}$. Therefore many known properties of the QAOA may be expected to hold for values of $d$ close to $N$ as well, as exemplified later.

Here, we restrict our attention to SK models~\cite{sherrington1975solvable}, which have quadratic cost functions of the form of Eq.~\eqref{eq:cost} with fully connected random weights $w_{ij}$ precisely defined later. Optimization problems expressed as fully connected graphs are ubiquitous in statistical mechanics and computer science~\cite{lucas2014ising}.

\section{Variational ansatz}
The QAOA ansatz for extremizing $C$ is $\ket{\psi_p} = e^{-i\beta_p\hat{\mathcal{H}}_x}e^{-i\gamma_p \hat{\mathcal{H}}} \ket{\psi_{p-1}}$~\cite{farhi2014quantum}. A natural way to generalize this ansatz for qubit-efficient encoding is to replace $\hat{\mathcal{H}}$ with $\hat{\mathcal{H}}[\psi_{p-1}]$. Here, $\hat{\mathcal{H}}_x = \sum_{i=1}^q \hat X_i$ is the transverse field mixer Hamiltonian, and $\hat{\mathcal{H}}[\psi_{p-1}]$ is the layer-dependent cost Hamiltonian, which depends on the wave function $\ket{\psi_{p-1}}$ in the previous layer. 

The ansatz is drawn in Fig.~\ref{fig:encoding+ansatz}(b). Determining the cost Hamiltonian in the $p^{\rm th}$ layer, $\hat{\mathcal{H}}[\psi_{p-1}]$, requires to first know $\bar{z}$ and $\braket{\psi_{p-1} \vert \hat P_\ell \vert \psi_{p-1}}$, which can be obtained from projective measurements after $p-1$ layers. Circuit implementation of $\exp(-i\gamma \hat{\mathcal{H}})$ is discussed in detail in Appendix~\ref{sec: circuit implementation}. Below, we exemplify this ansatz using an SK problem instance, and argue for a slight modification to the ansatz.

\subsection{Detailed illustration with an example}\label{subsec:example problem}
\begin{figure}[t!]\centering
\includegraphics[width=1.0\columnwidth]{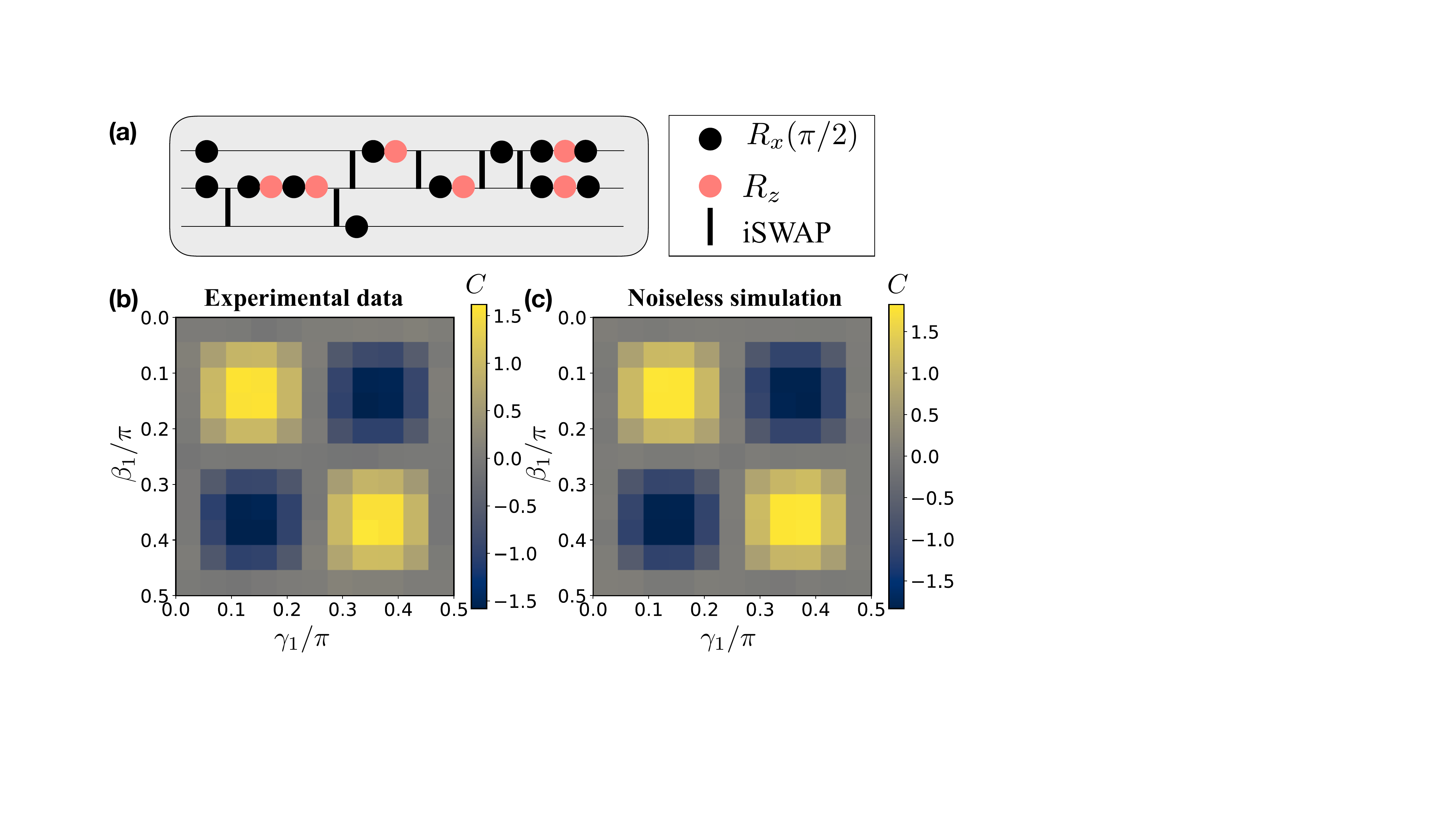}
\caption{(a) Schematic of quantum circuit that implements a $p=1$ layer of the ansatz, for an $N=4$ SK model instance at $d=2$. Orange and black dots are $R_x$ and $R_z$ gates, and vertical lines are iSWAPs (see Appendix~\ref{subsec: circuit implementation special case}). (b-c) Cost landscape for the SK problem instance in Eq.~\eqref{eq:example} versus $\beta_1$ and $\gamma_1$, measured from experiments on Rigetti's Ankaa\textsuperscript{TM}-9Q-3 chip, and numerically computed from an ideal simulation of the circuit, with 10000 shots.}
\label{fig:experimental_result}
\end{figure}
We consider an SK model problem with $N=4$ variables, where the weights $w_{ij}$ are chosen randomly from $\{\pm 1\}$. To illustrate the ansatz, we take the $d=2$ encoding as an example, which encodes the problem onto three qubits. 

Figure~\ref{fig:experimental_result}(a) schematically shows the structure of the circuit required to implement the ansatz at $p=1$, expressed in terms of quantum gates native to Rigetti's Ankaa\textsuperscript{TM}-9Q-3 superconducting quantum chip. To enable compilation of the same circuit into different gates suitable for other architectures, Appendix~\ref{subsec: circuit implementation special case} presents the logical form of the phase separator in terms of $ZZ$ and controlled-$ZZ$ rotations.

To implement the ansatz $\ket{\psi_{p=1}} = e^{-i\beta\hat{\mathcal{H}}_x}e^{-i\gamma\hat{\mathcal{H}}[\psi_0]}\ket{\psi_0}$, we first need to determine $\hat{\mathcal{H}}[\psi_0]$, where $\ket{\psi_0}$ is the equal superposition state of three qubits. Measurements of $\ket{\psi_0}$ would give
\begin{align}\label{eq:H special case}
& \bar{z}_1 = \bar{z}_2 = \bar{z}_3 = \bar{z}_4 = 0, \nonumber\\
& \braket{\psi_0 \vert \hat P_{\ell=0} \vert \psi_0} = \braket{\psi_0 \vert \hat P_{\ell=1} \vert \psi_0} = 1/2, \nonumber\\
\Rightarrow & \hat{\mathcal{H}}[\psi_0] = 2(w_{12} \hat P_{\ell=0}\hat Z_{d_1}\hat Z_{d_2} + w_{34} \hat P_{\ell=1}\hat Z_{d_1}\hat Z_{d_2}).
\end{align}

An example instantiation yielded the following weights,
\begin{equation}\label{eq:example}
w = \left( \begin{array}{cccc}
0 & 1 & -1 & 1\\
1 & 0 & -1 & -1\\
-1 & -1 & 0 & 1\\
1 & -1 & 1 & 0
\end{array}\right).
\end{equation}
The SK model objective function is thus
\begin{equation}
C(\boldsymbol{z}) = z_1 z_2 - z_1 z_3 + z_1 z_4 - z_2 z_3 - z_2 z_4 + z_3 z_4.
\end{equation}
It can be verified, e.g. by explicitly enumerating all the 16 bit strings for the four variables, that the ground-state solutions for this example are $\boldsymbol{z}^* = (1,-1,1,-1)$ and $-\boldsymbol{z}^* = (-1,1,-1,1)$, with the ground-state cost $C^* = -4$. The solutions can be represented by the respective wave functions $\ket{\phi(\boldsymbol{z}^*)} = (\ket{0}_{\textrm{label}}\otimes\ket{01}_{\textrm{data}} + \ket{1}_{\textrm{label}}\otimes\ket{01}_{\textrm{data}})/\sqrt{2}$ and $\ket{\phi(-\boldsymbol{z}^*)} = (\ket{0}_{\textrm{label}}\otimes\ket{10}_{\textrm{data}} + \ket{1}_{\textrm{label}}\otimes\ket{10}_{\textrm{data}})/\sqrt{2}$, and the ultimate goal of the ansatz is to prepare one of these wave functions.

Figures~\ref{fig:experimental_result}(b-c) plot the cost yielded by the ansatz versus $\beta_1$ and $\gamma_1$, with results measured on Rigetti's Ankaa\textsuperscript{TM}-9Q-3 in (b), and numerically obtained from a noiseless simulator in (c). A sample comparison between numerical and experimental data at specific ansatz parameters is shown in Appendix~\ref{sec: bit strings}. We find extremely good agreement between the cost landscapes, with a Pearson correlation coefficient of 0.997. There is a small $\sim15\%$ scale reduction of the cost on Ankaa\textsuperscript{TM}-9Q-3, which is attributable to hardware noise. For completeness, we also note the expectation value of the cost obtained from an explicit calculation,
\begin{align}
C[\psi_{p=1}] =& (w_{12}-w_{34})\sin2\beta_1 \sin(2\gamma_1(w_{12}-w_{34})) \nonumber\\ 
& (\cos 4\beta_1 + 2\cos^2 2\beta_1\cos(2\gamma_1(w_{12}+w_{34}))) \nonumber\\
& + \sin2\beta_1\cos2\beta_1(w_{12}+w_{34})(\sin(4\gamma_1w_{12}) 
\nonumber\\ & +\sin(4\gamma_1w_{34})).
\end{align}
For the specific example in Eq.~\eqref{eq:example}, the noiseless cost is $C[\psi_{p=1}] = 2\sin4\beta_1 \sin4\gamma_1$.

The parameters that minimize the cost at $p=1$ are $(\beta_1,\gamma_1)=\left(\frac{3\pi}{8},\frac{\pi}{8}\right)$, and the wave function $\ket{\psi_{p=1}}$ is 
\begin{align}
\ket{\psi_{p=1}} = & \frac{ (\ket{0}_{\textrm{label}}+\ket{1}_{\textrm{label}})\otimes(\ket{01}_{\textrm{data}} +\ket{10}_{\textrm{data}} )}{2} \nonumber\\
=& \frac{\ket{\phi(\boldsymbol{z}^*)}+\ket{\phi(-\boldsymbol{z}^*)}}{\sqrt{2}}.
\end{align}
The wave function is an equal superposition of $\ket{\phi(\boldsymbol{z}^*)}$ and $\ket{\phi(-\boldsymbol{z}^*)}$ due to $\mathbb{Z}_2$ symmetry of the ansatz, i.e. $\bar{z}_i=0\ \forall i$.
Notably, the $p=1$ ansatz does not produce the ground-state cost, and \textit{cannot} produce the ground-state cost at any $p$ due to the $\mathbb{Z}_2$ symmetry. We note that this is unique to the qubit-efficient encoding, and does not occur at $d=N$~\footnote{It is a unique feature of the qubit-efficient encoding that a \textit{superposition} of ground-state solutions does not have the same cost $C$ as the ground-state cost $C^*$, whereas the cost for the superposition would be equal to the ground-state cost $C^*$ in the $d=N$ case. This is because $\bar{z}_i=0$, therefore several terms in the cost at $d<N$ are 0. Therefore, explicitly breaking the $\mathbb{Z}_2$ symmetry in the ansatz is necessary.}.

\subsection{Modification to the ansatz}
As shown in the example above, breaking the $\mathbb{Z}_2$ symmetry of the ansatz is needed to produce the ground-state cost. Therefore, we modify the ansatz to
\begin{equation}\label{eq:ansatz}
 \ket{\psi_p} = e^{-i\beta_p\hat{\mathcal{H}}_x}e^{-i\gamma'_p \hat{\mathcal{H}}_z}e^{-i\gamma_p \hat{\mathcal{H}}[\psi_{p-1}]} \ket{\psi_{p-1}},
\end{equation}
where $\hat{\mathcal{H}}_z = \sum_{i\in\textrm{data}} \hat Z_i$ describes one-body bias terms.

\section{Variational optimization} 
\begin{figure}[t!]\centering
\includegraphics[width=0.8\columnwidth]{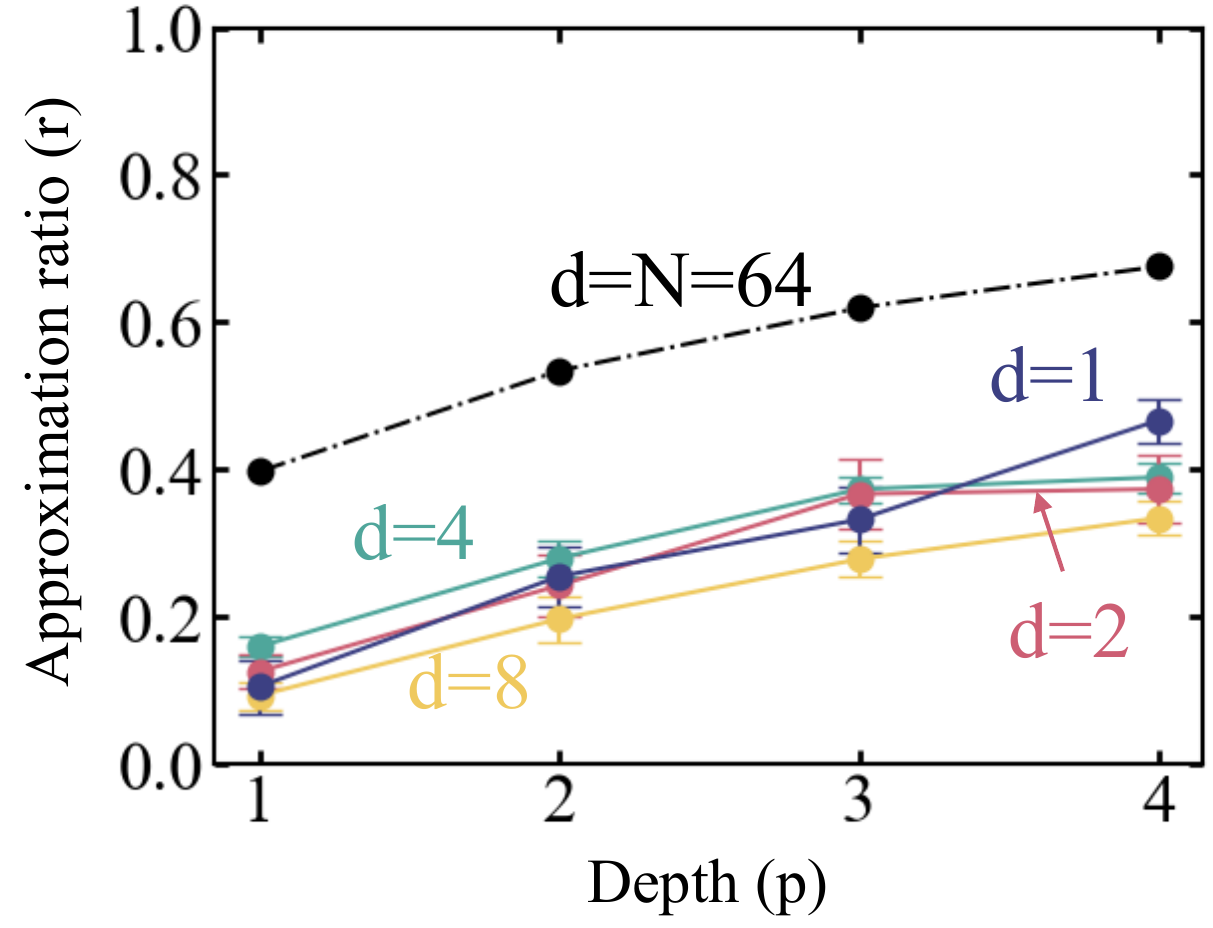}
\caption{Average approximation ratios produced by the optimal circuits in Fig.~\ref{fig:encoding+ansatz}(b), for ensemble of five Sherrington-Kirkpatrick spin glass model problems with $N=64$ variables and $w_{ij}\in\{\pm1\}$, at various values of $d$.}
\label{fig:result}
\end{figure}
We illustrate the performance of the symmetry-broken ansatz using an ensemble of SK problem instances at $N=64$ and $128$ with $w_{ij}\in\{\pm1\}$. We optimize the $(\beta_i,\gamma_i,\gamma'_i)$ parameters in the ansatz with a classical optimizer that seeks to minimize $C[\psi_p]$. We use a basin-hopping optimizer~\cite{wales1997global} with L-BFGS-B as the local optimization method~\cite{liu1989limited}. We empirically set the basin-hopping optimizer to iterate over $21p$ basins during the optimization, where each basin is a seed for a local search by the L-BFGS-B optimizer. We capped the total number of L-BFGS-B iterations in each basin at 1000. To benchmark the cost produced by variational optimization, we compare it against the ground-state cost $C^*$ found with a classical heuristic algorithm based on tabu search~\cite{glover2010diversification} implemented in MQLib~\cite{DunningEtAl2018}.

Figure~\ref{fig:result} plots the optimized approximation ratios, $r=C/C^*$, averaged over an ensemble of five SK problems with $N=64$ variables, using different values of $d<N$. The results at $d<N$ are obtained from a noiseless state vector simulator. The dashed line at $d=N$ is plotted using values predicted in Ref.~\cite{farhi2022quantum} for large $N$. Our ansatz at $(d,p)=(2,3)$ and $(4,3)$ produces results comparable to the ansatz at $(d,p)=(N,1)$, despite having up to 9 and 8 times fewer qubits, respectively. The ansatz at $d=1,p=4$ produces results better than the ansatz at $d=N,p=2$.

In addition to the quality of the solution produced by an algorithm, the time taken by the algorithm is also an important factor. A significant contributor to the total time of variational quantum algorithms is the time taken to find the optimal parameters.
Figure~\ref{fig:result} shows no clear distinction in the quality of solution produced by different $d$ at the same $p$. But we argue next that larger values of $d$ may require lesser parameter search time due to clustering of optimal parameters for SK problem instances.

\subsection{Concentration of parameters} 
\begin{figure*}[t!]\centering
\includegraphics[width=2.0\columnwidth]{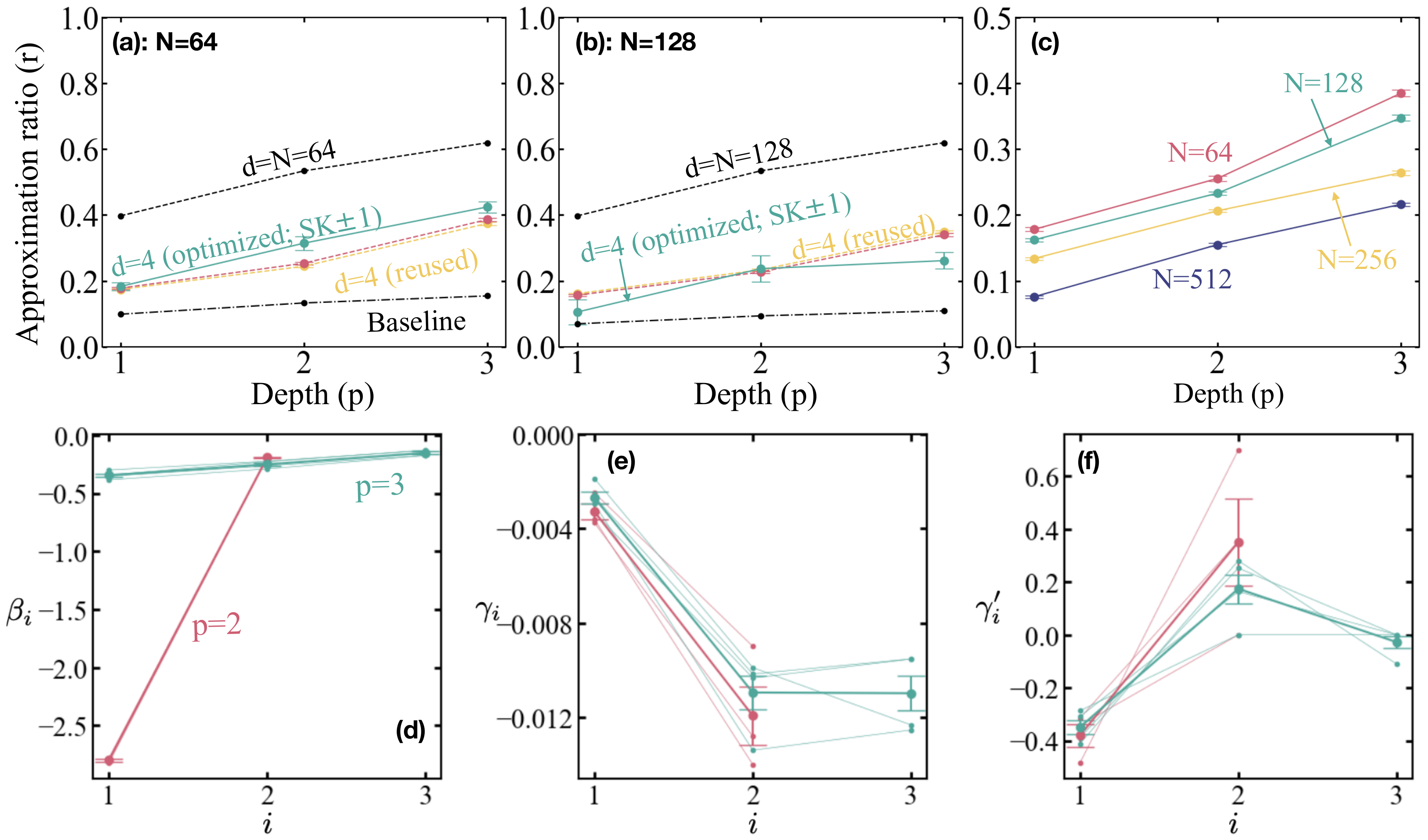}
\caption{Evidence for parameter concentration in the qubit-efficient ansatz for the SK model. (a-c) Average approximation ratios for the SK model versus depth $p$. (a-b) plot average approximation ratios for optimized parameters for five SK problem instances (teal), and for 100 SK problem instances where the same parameters were used for all the instances (yellow and magenta). Instances in yellow have $w_{ij}=\pm1$ while magenta instances have $w_{ij}\sim \mathcal{N}(0,1)$. The parameters for the yellow and magenta curves in (b) are related to those of (a) by a trivial rescaling factor (see text). (c) plots average approximation ratios for 100 problem instances at different sizes, solved by reusing and rescaling parameters from (a). (d-f) Optimal parameters returned by the classical optimizer for the five SK problem instances with $N=64$ variables in Fig.~\ref{fig:result}, at $d=4$. The thin curves plot the instance-by-instance optimal parameters, and the thick curves plot their mean values. We find the curves are clustered together.}
\label{fig:concentration}
\end{figure*}

\begin{figure}[!t]
\includegraphics[width=0.7\columnwidth]{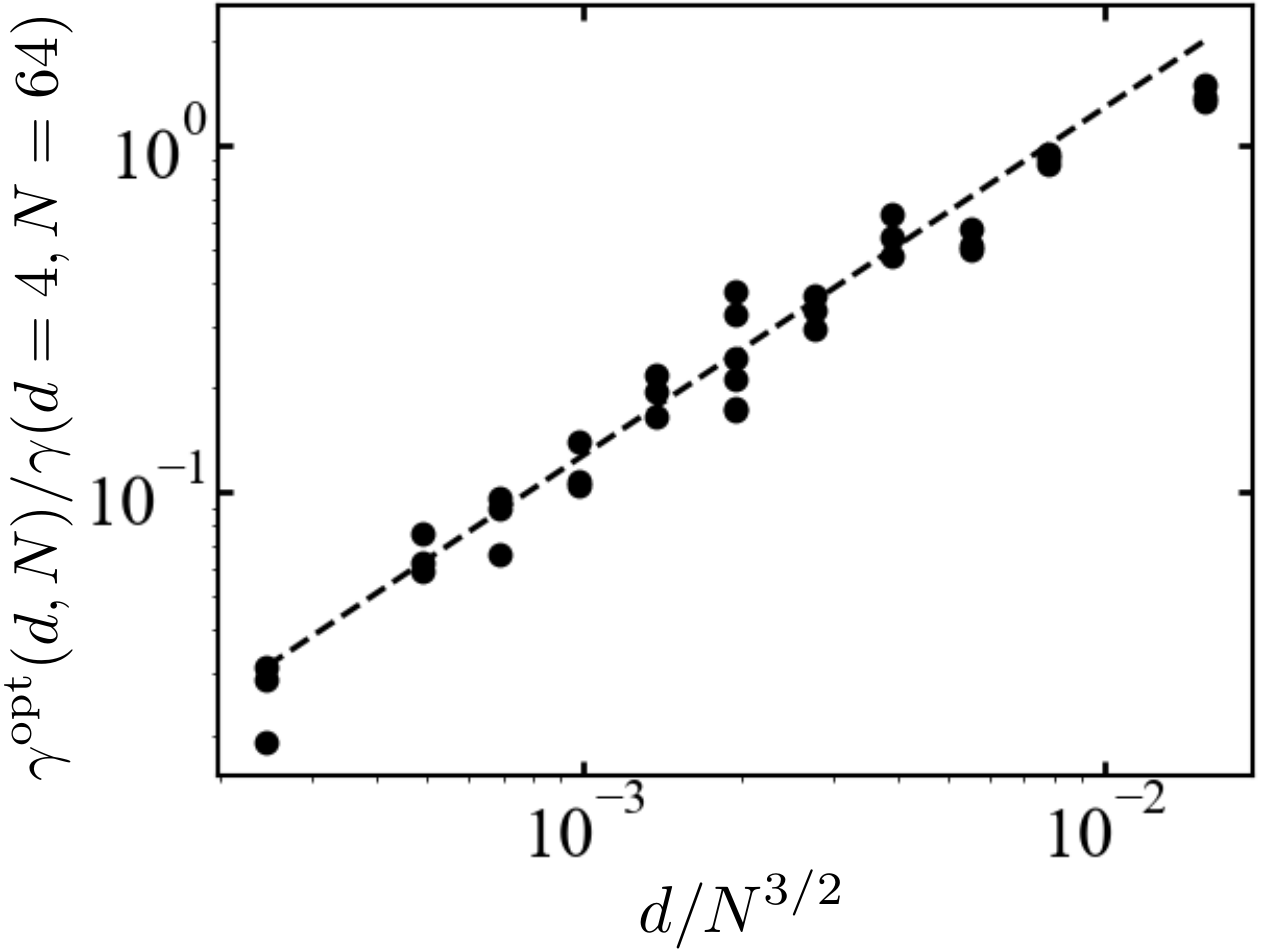}
\caption{Optimal ratios $\gamma^{\rm opt}(N,d; p,i)/\gamma(N=64,d=64; p,i)$ scale as $d/N^{3/2}$. The points include $1\leq p\leq 3$, $N\in\{64,128,256\}, d\in\{1,2,4,8\}$, with one SK problem instance at each $N$ and $d$. Dashed line plots $d/N^{3/2}$ as guide to the eye.}
\label{fig:gamma_scaling}
\end{figure}

Previous works have shown empirical and analytical evidence for clustering of optimal parameters in the QAOA ($d=N$) for some classes of problems, including SK problems~\cite{zhou2020quantum, farhi2022quantum, basso2022quantum, wurtz2021fixed, galda2023similarity, brady2021optimal}. It is interesting to ask whether a similar property holds for the qubit-efficient ansatz.

In this context, we note that QAOA's optimal parameters and approximation ratios for the typical SK problem instance have been tabulated up to $p=12$~\cite{farhi2022quantum}. Our cost function $C$ is related to the cost $\tilde{C}$ in Ref.~\cite{farhi2022quantum} via $C = \sqrt{N}\tilde{C}$, and our parameters $(\beta, \gamma)$ at $d=N$ are related to parameters $(\tilde{\beta}, \tilde{\gamma})$ in Ref.~\cite{farhi2022quantum} via $\beta = \tilde{\beta}$ and $\gamma = \tilde{\gamma}/\sqrt{N}$. The optimal value of $\gamma$ scales as $1/\sqrt{N}$.

Figure~\ref{fig:concentration} demonstrates empirical evidence for parameter concentration for SK problems in the qubit-efficient ansatz. Figures~\ref{fig:concentration}(d-f) plot the optimal parameters found by the classical optimizer for the five problem instances in Fig.~\ref{fig:result}, up to $p=3$ at $d=4$ and $N=64$.
We implemented the ansatz for solving 100 SK problem instances, using the optimized parameters for one SK problem instance from Fig.~\ref{fig:result}. The yellow dashed line in Fig.~\ref{fig:concentration}(a) plots the mean $r$ vs $p$ for 100 SK problem instances with $w_{ij}=\pm1$, and shows that the mean $r$ for 100 problem instances closely follows the optimized $r$ for five problem instances (solid teal curve).

Figure~\ref{fig:concentration}(b) demonstrates another useful property that near-optimal parameters for $N>64$ can be derived from those at $N=64$. We solve SK problems with $N=128$, by reusing the same parameters from the $N=64$ case in (a) with a trivial rescaling of $\gamma$ (see below), and still find approximation ratios close to the optimized $r$ for five problem instances (solid teal curve).

To explain the occurrence of parameter concentration at $d<N$, we hypothesize that the optimal coefficient $\gamma$ multiplying $\hat Z_i\hat Z_j$ in the qubit-efficient ansatz scales $\propto 1/\sqrt{N}$, just like in the $d=N$ case. We validate this hypothesis as follows. Ising terms such as $\hat Z_i\hat Z_j$ appear in the ansatz via $\gamma \frac{\hat P_\ell \hat Z_{d_i}\hat Z_{d_j}}{\braket{\hat P_\ell}}$. Here, $\braket{\hat P_\ell}=d/N$ in the equal superposition state which is the $p=0$ state of the ansatz, although it will vary with $p$. Setting $\gamma\frac{N}{d} \propto 1/\sqrt{N}$ gives $\gamma(N,d) \propto 1/N^{3/2}$. This is the rescaling we applied when we used optimal parameters from $N=64$ to solve SK problems instances with $N=128$ in Fig.~\ref{fig:concentration}(b), i.e. $
\gamma^*(N=128,d=4) = \gamma^*(N=64,d=4)/2^{3/2}$. We perform the same rescaling and calculate the approximation ratios for an ensemble of 100 SK problem instances with $N=256,512$ in Fig.~\ref{fig:concentration}(c), and find the approximation ratio increases with $p$.

We validate the scaling of $\gamma$ with the following optimization procedure. For an ensemble of SK problems at different sizes and $d$, we fix the values of $\beta(N,d;p,i)$ and $\gamma'(N,d;p,i)$ from one instance at $N=64,d=4$, and vary the values of $\gamma(N,d;p,i)$, where $1 \leq i \leq p$. We assume that $\gamma(N,d;p,i)/\gamma(N=64,d=4;p,i) = \theta(N,d;p)$ is a constant, and optimize over $\theta(N,d;p)$. Figure~\ref{fig:gamma_scaling} finds that $\theta^{\rm opt}(N,d;p) \propto d/N^{3/2}$, as predicted by the above argument.

Furthermore, we recall that QAOA's optimal values of $(\beta, \gamma)$ and $C$ up to $p=11$ are the same for two classes of SK models --- $w_{ij}\in\{\pm1\}$, and $w_{ij}\sim\mathcal{N}(0,1)$.
This property seems to hold for $d<N$ as well. The magenta dashed line in Fig.~\ref{fig:concentration}(a), which plots the mean approximation ratio for problems with normally distributed weights, $w_{ij}\sim\mathcal{N}(0,1)$, demonstrates that the same parameters obtained from the case with $w_{ij}\in\{\pm1\}$ can also be used to solve the ensemble with $w_{ij}\sim\mathcal{N}(0,1)$. Finally, our ansatz beats a baseline performance that comes from a naive problem decomposition method, explained in Appendix~\ref{sec: baseline}.

\section{Adaptive ansatz}

\begin{figure}[!t]\centering
\includegraphics[width=0.8\columnwidth]{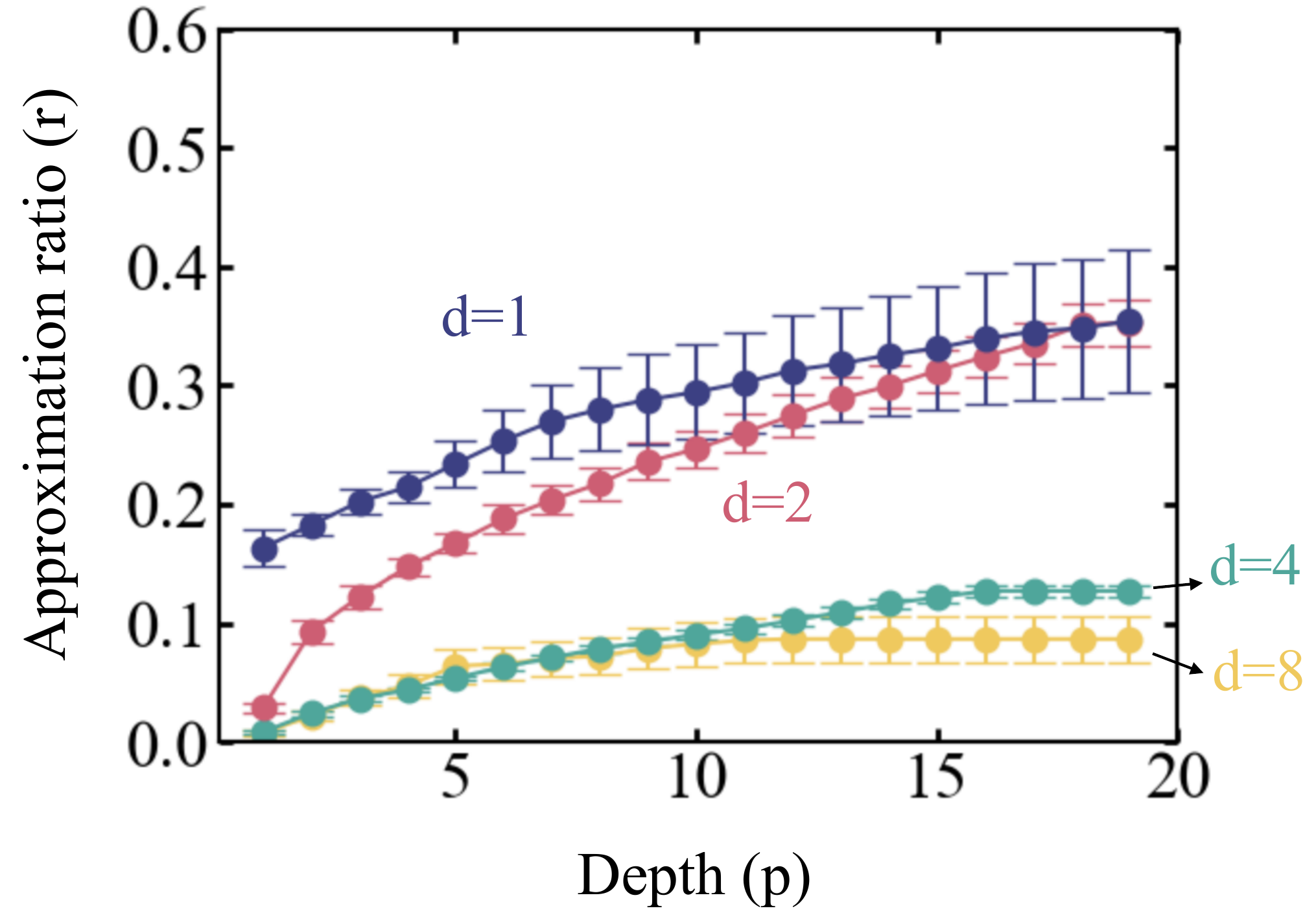}
\caption{Average approximation ratios produced by the qubit-efficient adaptive quantum approximate optimization, for ensemble of five Sherrington-Kirkpatrick spin-glass-model problems with $N=64$ variables and $w_{ij}\in\{\pm1\}$, at various values of $d$.}
\label{fig:adapt}
\end{figure}

Earlier works on variants of QAOA have provided some numerical evidence that they can perform better than QAOA. These variants may include counterdiabatic terms in the phase separator, or adaptively chosen problem-tailored mixers, with the pool of mixers also inspired by the concept of shortcuts to adiabaticity. We note that the ADAPT-QAOA~\cite{zhu2022adaptive}, which is one of the algorithms that adaptively chooses the mixers, has some similarity to the qubit-efficient algorithm we have proposed -- it builds the ansatz layer by layer by choosing the best mixer in each layer, similar to how we calculate the phase separator in each layer.

Here, we perform a small-scale preliminary study of an adaptive qubit-efficient variational ansatz, which combines ideas from the concept of \textit{mixer operator pool} from the ADAPT-QAOA, and phase separator in the qubit-efficient QAOA in the previous sections.

The algorithm starts with all the qubits initialized in $\ket{+}$, as usual. We construct an operator pool for the mixers, which includes $\sum_i X_i$, but also other mixers. For this study, we chose a pool of mixers identical to that in Ref.~\cite{zhu2022adaptive}. Explicitly, the operator pool is
\begin{align}
S = &\{ \sum_i X_i, \sum_i Y_i\}\ \bigcup\ \bigcup_{i=1}^{q}\{X_i, Y_i\} \bigcup \ \nonumber\\
& \bigcup_{P,P'\in \{X,Y,Z\}}\bigcup_{i=1}^{q}\bigcup_{j=1}^{q} \{ P_i P'_j \}
\end{align}
where $q = d + \log_2(N/d)$ is the number of qubits. 

In the $p$th layer, we calculate the gradient $\delta_p(\hat{s})$ with respect to each mixer $\hat{s} \in S$ as
\begin{align}
& \ket{\psi_p(\beta; \hat{s})} = e^{-i \beta \hat{s}} e^{-i\gamma_0 \hat{H}[\psi_{p-1}]} \ket{\psi_{p-1}} \nonumber\\
&\delta_p(\hat{s}) = \lim_{\beta \rightarrow 0} \frac{ C[\psi_p(\beta; \hat{s})] - C[\psi_p(0; \hat{s}] }{ \beta },
\end{align}
where $\gamma_0$ is some predefined value. Here, we arbitrarily choose $\gamma_0 = \pi/\sqrt{N}$. We recall that the cost $C[\psi]$ for the wave function $\psi$ is $C[\psi] = \braket{\psi \vert \hat{H}[\psi] \vert \psi}$. The mixer $\hat{s}$ for the $p$th layer is that with the largest gradient, namely $|\delta_p(\hat{s})|$.

After choosing the mixer, we optimize the $\beta_p$ and $\gamma_p$ parameters for that layer, using a grid search, fixing all parameters and mixers up to $(p-1)$ layers. This layer-by-layer greedy construction of the ansatz and search of parameters is guaranteed to yield a nondecreasing cost, but not necessarily the optimal cost at $p$ layers. We discuss potential improvements to this strategy later.

Figure~\ref{fig:adapt} shows the results of this small-scale study for the same ensemble of SK problem instances and same values of $d$ as Fig.~\ref{fig:result}. We find that the average approximation ratio steadily increases with $p$, and then saturates at some $p$, presumably because the algorithm has reached a local minimum. At $d=8$, the approximation ratio saturates at $p=10$, and the approximation ratio does not increase further after the tenth layer. The approximation ratio saturates at $p=15$ for $d=4$, and at $p=17$ for $d=2$. The approximation ratio has not yet saturated for $d=1$, but has a small gradient at $p=19$. We also note that these approximation ratios are smaller than those found for qubit-efficient QAOA in Fig.~\ref{fig:result}.

Improvements to our adaptive algorithm may increase the approximation ratios in Fig.~\ref{fig:adapt}. Some improvements in the concept of the traditional encoding ($d=N$) have already been suggested~\cite{yanakiev2024dynamic}. Further, instead of greedily optimizing $(\beta_p, \gamma_p)$ while fixing all parameters in previous layers, one improvement strategy could globally optimize $\{\beta_i, \gamma_i\}_{i=1}^p$ after fixing the mixers up to $p$ layers. Finally, while we chose the same operator pool as Ref.~\cite{zhu2022adaptive} but a better mixer pool may be necessary for qubit-efficient encoding. The mixers in the operator pool are inspired by the Lie algebra of $\hat{H}$ and $\sum_i X_i$. Since $\hat{H}$ has the form in Eq.~\eqref{eq:H} for qubit-efficient encoding, the mixer pool may also need to be state-dependent. We leave these explorations to a future study.

\section{Summary}
We showed that a qubit-efficient QAOA-inspired algorithm for solving SK spin-glass problems, which maps many variables to each qubit in a register, achieves a similar approximation ratio to QAOA, while using substantially fewer qubits. We demonstrated a proof-of-principle experimental benchmark of the algorithm for solving a small SK model problem on Rigetti's Ankaa\textsuperscript{TM}-9Q-3 superconducting chip. We derived a performance guarantee for our ansatz, and showed evidence of parameter clustering at $d=4$, which could significantly alleviate classical resources for parameter optimization. We did not find parameter concentration at $d<4$, which could indicate one of the difficulties in implementing the algorithm with $O(\log_2N)$ qubits, and therefore a preference for larger $d$. However, this preference for more qubits should be weighed against the increased hardware noise in the circuit when more qubits are present. Moreover, one should consider the fact that the solution bit string would be stored as an entangled final wave function, which may be less robust to noise when more qubits are used. In this context, error mitigation and suppression techniques~\cite{cai2023quantum, lidar2014review, suter2016colloquium} could be useful to improve the results of the qubit-efficient algorithm. Thus, investigating the optimal value of $d$ on real hardware, with error mitigation and suppression techniques, would be an interesting question to explore in the future.

Our algorithm can be straightforwardly extended to other classes of combinatorial problems, and is expected to have a trainable QAOA-like ansatz that may have parameter clustering or at least good heuristics for seed parameters, and lesser overheads than standard problem decomposition methods~\cite{angone2023hybrid, bach2024mlqaoa, liu2022hybrid, bravyi2016trading,peng2020simulating,tang2021cutqc,decross2023qubit,kim2023evidence, bechtold2023investigating,dupont2025benchmarking}. 
The qubit-efficient scheme may also be applied to other use cases such as quantum machine learning, and is likely to be useful in the near future as we enter the era of small-scale fault-tolerant quantum computers~\cite{bluvstein2024logical, google2023suppressing, google2025quantum, da2024demonstration, krinner2022realizing}.

\begin{acknowledgments}
We thank Stuart Hadfield, Mark Hodson, Marco Paini, Alex Place, Matt Reagor, Graham Reid, and Davide Venturelli for valuable discussions.
This work is supported by the Defense Advanced Research Projects Agency (DARPA) under Agreement No. HR-0011-23-3-0015.
This research used resources of the National Energy Research Scientific Computing Center, a DOE Office of Science User Facility supported by the Office of Science of the U.S. Department of Energy under Contract No. DE-AC02-05CH11231 using NERSC awards DDR-ERCAP0024427 and ASCR-ERCAP0028951. The views, opinions, and/or findings expressed are those of the authors and should not be interpreted as representing the official views or policies of the Department of Defense or the U.S. Government.
\end{acknowledgments}

\appendix

\section{Qubit-efficient encoding}
In Sec.~\ref{sec: encoding}, we described the qubit-efficient encoding of $N$ classical variables [Eq.~\eqref{eq:dense-encoding}], and how to estimate $\bar{z}_i$ and $\overline{z_iz_j}$ [Eqs.~\eqref{eq:zbar} and~\eqref{eq:zzbar}]. Here, we exemplify these using the $(N=4,\ d=2)$ case as an example.

Encoding $N=4$ variables using $d=2$ requires three qubits. The most generic three-qubit wave function is
\begin{equation}
\ket{\psi} = c_0\ket{000} + c_1\ket{001} + \cdots + c_7\ket{111}
\end{equation}
with $\sum_{i=0}^7 |c_i|^2 = 1$. Denoting the most significant bit as the label bit, projectively measuring this wave function gives the following measurements with respective probabilities
\begin{equation}
\begin{array}{c|c|c}
\textrm{Variables} & \textrm{Values} & \textrm{Probability}\\
\hline
(z_0, z_1) & (1,1) & |c_0|^2\\
(z_0, z_1) & (-1,1) & |c_1|^2\\
(z_0, z_1) & (1,-1) & |c_2|^2\\
(z_0, z_1) & (-1,-1) & |c_3|^2\\
(z_2, z_3) & (1,1) & |c_4|^2\\
(z_2, z_3) & (-1,1) & |c_5|^2\\
(z_2, z_3) & (1,-1) & |c_6|^2\\
(z_2, z_3) & (-1,-1) & |c_7|^2\\
\hline
\end{array}
\end{equation}
In an experiment with $N_s$ shots, let $N_s^{(0)}$ denote the number of shots that yield the values of $(z_0, z_1)$, and $N_s^{(1)}$ the number of shots that yield the values of $(z_2, z_3)$. The expected values of $N_s^{(0, 1)}$ are $E[N_s^{(0)}] = N_s\sum_{i=0}^3 |c_i|^2$ and $E[N_s^{(1)}] = N_s\sum_{i=4}^7 |c_i|^2$. The estimate for, e.g., $\overline{z_0 z_1}$, is obtained from just the $N_s^{(0)}$ shots as
\begin{align}\label{eq:z0z1}
\overline{z_0 z_1} = &\frac{\sum_{i,j=\pm 1}ij N(z_0=i,z_1=j)}{N_s^{(0)}}\nonumber\\
 \simeq &\frac{ |c_0|^2-|c_1|^2-|c_2|^2+|c_3|^2 }{ |c_0|^2+|c_1|^2+|c_2|^2+|c_3|^2 }
\end{align}
where $N(z_0=i,z_1=j)$ is the number of times $(z_0,z_1)$ were measured as $(i,j)$. 
The expression in Eq.~\eqref{eq:z0z1} is equal to $\frac{\braket{\hat P_{\ell=0}\hat Z_1\hat Z_2}}{\braket{\hat P_{\ell=0}}}$. Other correlations and one-body terms can be similarly estimated as
\begin{align}
\overline{z_2 z_3} =& \frac{ |c_4|^2-|c_5|^2-|c_6|^2+|c_7|^2 }{ |c_4|^2+|c_5|^2+|c_6|^2+|c_7|^2 }\nonumber\\
\bar{z}_0 =& \frac{ |c_0|^2-|c_1|^2+|c_2|^2-|c_3|^2 }{ |c_0|^2+|c_1|^2+|c_2|^2+|c_3|^2 }\nonumber\\
\bar{z}_1 =& \frac{ |c_0|^2+|c_1|^2-|c_2|^2-|c_3|^2 }{ |c_0|^2+|c_1|^2+|c_2|^2+|c_3|^2 }\nonumber\\
\bar{z}_2 =& \frac{ |c_4|^2-|c_5|^2+|c_6|^2-|c_7|^2 }{ |c_4|^2+|c_5|^2+|c_6|^2+|c_7|^2 }\nonumber\\
\bar{z}_3 =& \frac{ |c_4|^2+|c_5|^2-|c_6|^2-|c_7|^2 }{ |c_4|^2+|c_5|^2+|c_6|^2+|c_7|^2 }
\end{align}
The estimate for the cost $C=\sum_{ij} w_{ij} z_i z_j$ is
\begin{align}
C = &w_{12}\overline{z_1z_2} + w_{34}\overline{z_3z_4} \nonumber\\
&+ w_{13}\bar{z}_1\bar{z}_3 + w_{14}\bar{z}_1\bar{z}_4 + w_{23}\bar{z}_2\bar{z}_3 + w_{24}\bar{z}_2\bar{z}_4.
\end{align}

\subsection{Case $d=N$}
In the $d=N$ case, there are no label bits, and each variable is mapped one to one with a data bit. This is the usual way variables are mapped to qubits. Correlations $z_i z_j$ are, e.g., estimated as $\overline{z_i z_j} = \braket{\hat Z_i \hat Z_j}$, and $C=\sum_{ij} w_{ij} z_i z_j$ is $C=\braket{\hat{\mathcal{H}}}$ with $\hat{\mathcal{H}} = \sum_{ij} w_{ij} \hat Z_i \hat Z_j$.

\subsection{Case $d=1$}
In the $d=1$ case, there are $\log_2N$ label bits, and all the variables are mapped to one data bit. All correlations $z_i z_j$ are estimated as $\overline{z_i z_j} = \bar{z}_i\times\bar{z}_j$.

\subsection{Case $d=2$}
The $d=2$ case, coincidentally, requires the same number of qubits as $d=1$. However, the mappings between qubits and variables are different. For $d=2$, there are $\log_2N-1$ label bits and $2$ data bits, unlike the $d=1$ case.

\section{Experimental data for the example problem in Fig.~\ref{fig:experimental_result}}\label{sec: bit strings}
\begin{figure}[!t]\centering
\includegraphics[width=0.9\columnwidth]{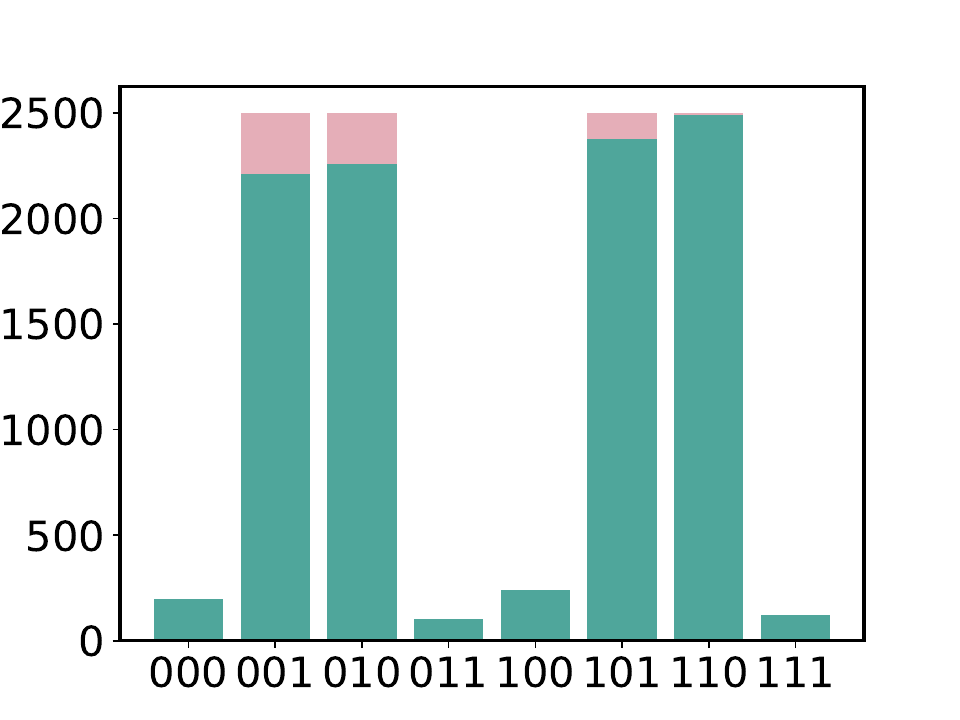}
\caption{Distribution of bit strings measured on Rigetti's Ankaa\textsuperscript{TM}-9Q-3 chip (teal), and predicted from the wave function (magenta), at the optimal angles of the ansatz for the example in Fig.~\ref{fig:experimental_result}.
}
\label{fig:bit strings}
\end{figure}
In Sec.~\ref{subsec:example problem}, we exemplified the variational ansatz using an SK problem instance at $N=4$. For that instance, the optimal ansatz parameters at $p=1$ are $(\beta_1,\gamma_1)=\left(\frac{3\pi}{8},\frac{\pi}{8}\right)$. Figure~\ref{fig:bit strings} plots the distribution of bit strings sampled from the wave function prepared on Rigetti's Ankaa\textsuperscript{TM}-9Q-3 chip (teal), and from a noiseless simulation of the ansatz (magenta). The noiseless cost at these parameters is $C[\psi_{p=1}] = -(w_{12}+w_{34}) = -2$, and the measured cost on Rigetti's Ankaa\textsuperscript{TM}-9Q-3 is $C = -1.736(1)$, which is a nearly $15\%$ reduction as mentioned in the main text.

\section{Entanglement in the qubit-efficient encoding}
Entanglement is an essential property of quantum devices and algorithms. A high degree of entanglement is the reason for nonsimulability and potential utility of quantum optimization. The growth of entanglement induced by the QAOA circuit has been previously studied~\cite{chen2022much, dupont2022entanglement, nakhl2024calibrating}, finding, for example, volume-law entanglement in the wave function between the initial and final state. The target state of QAOA is a product state with no entanglement.

In the qubit-efficient approach, however, the target state is not a product state. Instead, the bit-string solution $\boldsymbol{z}$ is stored as the entangled wave function $\ket{\phi(\boldsymbol{z})} = \sum_\ell \lambda_\ell \ket{\ell}\otimes\ket{\boldsymbol{z}_\ell}$ [Eq.~\eqref{eq:dense-encoding}]. $\boldsymbol{z}_\ell$ are length-$d$ subsets of $\boldsymbol{z}$, where the subscript $\ell$ runs from $0$ to $N/d-1$. Here, we quantify the entanglement present in the target wave function by calculating the entanglement entropy of the data qubits.

The reduced density matrix for the data qubits, obtained by tracing over the label qubits, is
\begin{equation}
\rho_{\rm data} = \sum\nolimits_{\ell=0}^{N/d-1} |\lambda_\ell|^2 \ket{\boldsymbol{z}_\ell}\bra{\boldsymbol{z}_\ell},
\end{equation}
and its von Neumann entropy is $S_{\rm data} = -{\rm Tr}(\rho_{\rm data}\log_2\rho_{\rm data})$.

\begin{figure}[!t]
\includegraphics[width=0.8\columnwidth]{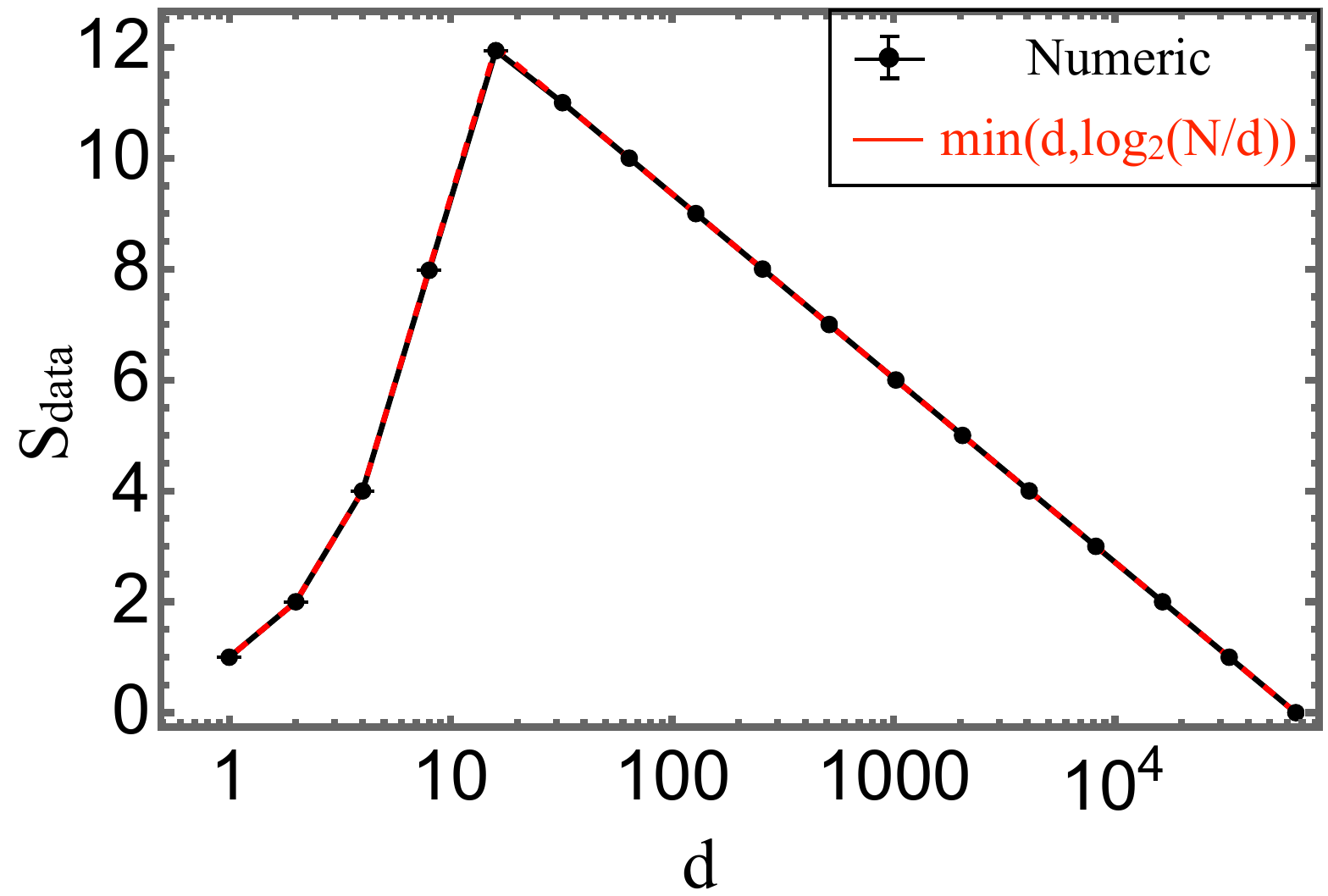}
\caption{Average entanglement entropy of the data qubits that store bit strings of length $N=2^{16}$. The data qubits have volume-law entanglement.}
\label{fig:supp_entropy}
\end{figure}

Figure~\ref{fig:supp_entropy} plots the average entanglement entropy, averaged over 100 random bit strings of length $N=2^{16}$, versus $d$, assuming wave functions with uniform $|\lambda_\ell|^2=d/N$. Two regimes emerge. When $d\lesssim 16$, the entanglement entropy increases as $d$, whereas for $d\gtrsim 16$, it decreases as $\log_2(N/d)$. This entropy is the smaller of the number of data and label qubits, indicating volume-law entanglement.

We can understand the two regimes as follows. When we encode a random bit string $\boldsymbol{z}$ at large $d$, i.e. $2^d \gg N/d$, the probability that any two $\boldsymbol{z}_\ell$ are equal is exponentially small in $d$; therefore, on average, $S_{\rm data} = -\sum_{\ell=0}^{N/d-1} |\lambda_\ell|^2 \log_2|\lambda_\ell|^2$. For uniform $|\lambda_\ell|^2 = d/N$, the entropy reduces to $S_{\rm data} = \log_2(N/d)$.

When $2^d \ll N/d$, there is a significant probability that there will be repetitions among the different $\boldsymbol{z}_\ell$. The average number of times any $\boldsymbol{z}_\ell$ would be repeated in $\boldsymbol{z}$ is $\frac{N/d}{2^d}$. Those repetitions must be grouped together in $\rho_{\rm data}$. For uniform $\lambda_\ell^2 = d/N$, this gives $\rho_{\rm data} \sim \sum_{\boldsymbol{z}_\ell=0}^{2^d-1} \frac{d}{N} \frac{N/d}{2^d} \ket{\boldsymbol{z}_\ell}\bra{\boldsymbol{z}_\ell}$. The entropy of this state is $S_{\rm data} = d$.

\section{Circuit implementation of the ansatz}\label{sec: circuit implementation}
\begin{figure*}[!t]
\includegraphics[width=1.0\textwidth]{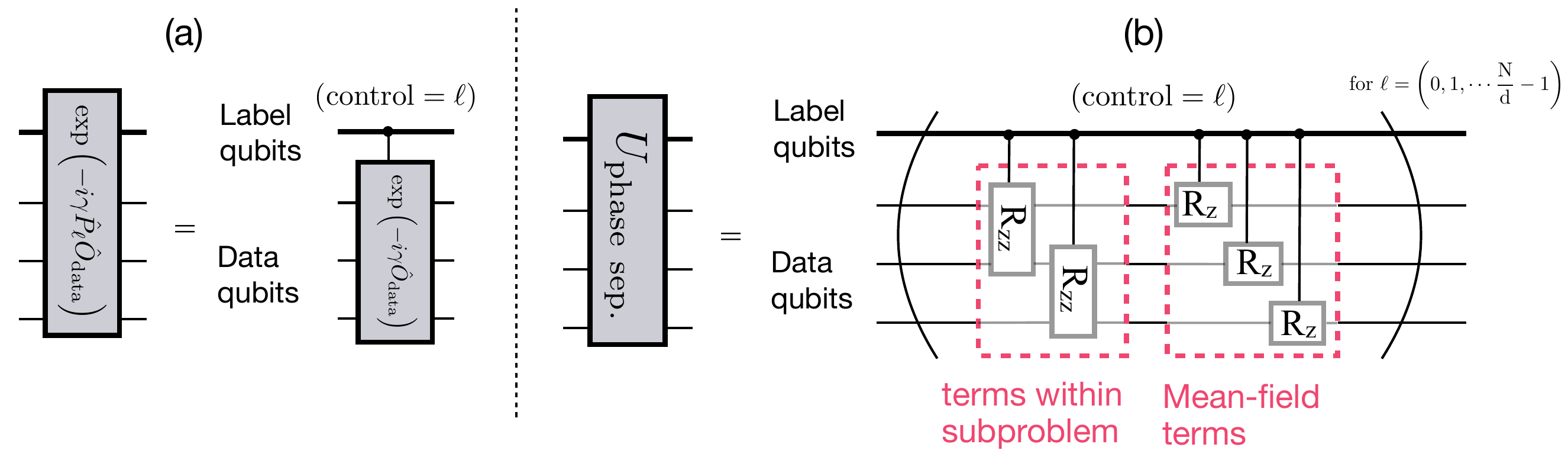}
\caption{Circuit implementation of the phase separator unitary in the ansatz [Eq.~\eqref{eq:H}]. The phase separator unitary is composed of terms of the form $\exp(i\gamma \hat P_\ell \hat O_{\rm data})$, which can be expressed as a controlled unitary as shown in (a). Here, $O_{\rm data}$ has either one-body $Z$ or two-body $ZZ$ terms. (b) The full phase separator is expressed as a product of terms in (a), with $\ell$ iterating from 0 to $N/d-1$.}
\label{fig:ansatz-implementation}
\end{figure*}

\begin{figure}[!t]
\includegraphics[width=0.8\columnwidth]{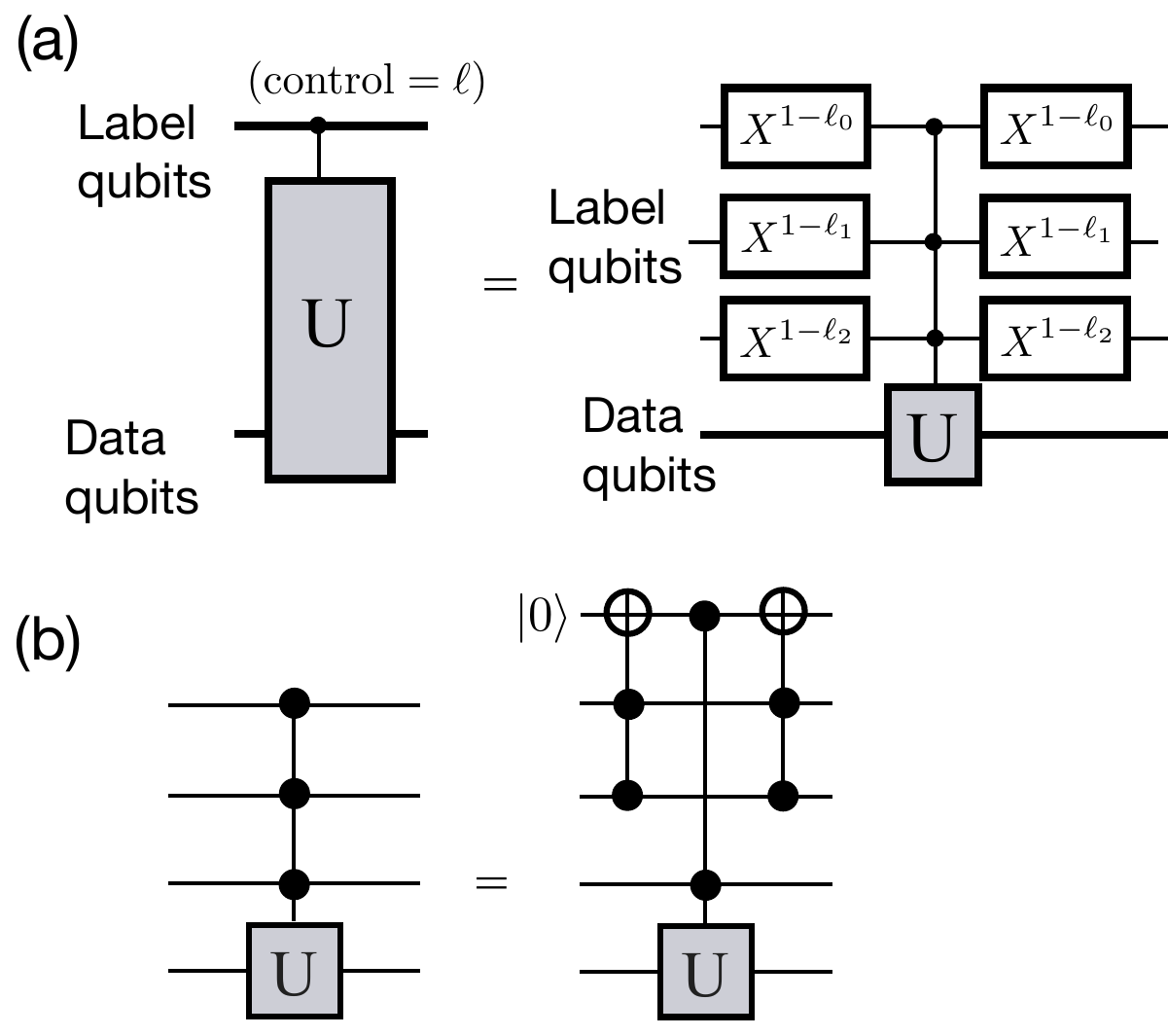}
\caption{Circuit identities for (a) implementing multicontrolled unitary, conditioned on the control bits having value $\ell$, in terms of a more standard multicontrolled unitary which is conditioned on the control bits having value $0$, and (b) recursively reducing controlled unitary with $m$ control bits to a controlled unitary with $(m-1)$ control bits and one ancilla bit.}
\label{fig:controlled unitary}
\end{figure}
\subsection{Phase separator unitary}
The phase separator Hamiltonian [Eq.~\eqref{eq:H}] at the $(p+1)$th layer is
\begin{align}
\hat{\mathcal{H}}[\psi_p] = & \sum_{\substack{i<j\\ \ell_i=\ell_j}} w_{ij} \frac{ \hat P_{\ell_i}\hat Z_{d_i}\hat Z_{d_j} }{\braket{\psi_p \vert \hat P_{\ell_i} \vert \psi_p}}
\nonumber\\
+ & \frac{1}{2}\sum_{\substack{i<j\\ \ell_i \neq \ell_j}} w_{ij} \left(\bar{z}_i \frac{ \hat P_{\ell_j}\hat Z_{d_j} }{\braket{\psi_p \vert \hat P_{\ell_j} \vert \psi_p}} 
+ \bar{z}_j \frac{ \hat P_{\ell_i}\hat Z_{d_i} }{\braket{\psi_p\vert \hat P_{\ell_i} \vert \psi_p}}\right).
\end{align}
The Hamiltonian is a sum of terms of the form of $\hat P_\ell O_{\rm data}(\ell)$, where $\ell$ summed from $0$ to $N/d-1$ refers to the label index, $\hat P_\ell$ is the projection operator on the label qubits, and $O_{\rm data}(\ell)$ is an operator that acts only on the data qubits. Since all the terms in $\hat{\mathcal{H}}[\psi_p]$ commute with each other, the unitary $\exp(-i\gamma_{p+1}\hat{\mathcal{H}}[\psi_p])$ is a product of $\exp\left(-i\gamma_{p+1} \hat P_\ell O_{\rm data}(\ell)\right)$.

To understand the circuit implementation of $\exp\left(-i\gamma_{p+1} \hat P_\ell O_{\rm data}(\ell)\right)$, let us examine how this unitary would act on the wave function $\ket{\psi_p}$. Let us write the Schmidt decomposition of $\ket{\psi_p}$ as
\begin{equation}
\ket{\psi_p} = \sum_\ell c_\ell \ket{\ell}_{\rm label} \ket{\psi_\ell}_{\rm data}.
\end{equation}
Then,
\begin{align}
\exp & \left(-i\gamma_{p+1} \hat P_{\ell'} \hat O_{\rm data}(\ell')\right)  \ket{\psi_p} = \sum_\ell c_\ell \ket{\ell}_{\rm label} \nonumber \\ 
\times & \left( e^{-i \gamma_{p+1}\delta_{\ell \ell'} \hat O_{\rm data}(\ell)} \ket{\psi_\ell}_{\rm data}\right).
\end{align}
Thus, we see that $\exp\left(-i\gamma_{p+1} \hat P_{\ell'} \hat O_{\rm data}(\ell')\right)$ is a controlled unitary operator, with the control bits being the label bits, applied only when the label bits have the value $\ell'$, the target bits are the data bits, and the target unitary is $\hat O_{\rm data}(\ell')$. This is illustrated in Fig.~\ref{fig:ansatz-implementation}(a).

Figure~\ref{fig:ansatz-implementation}(b) shows the full implementation of the phase separator. The circuit implementation has a sequence of multicontrolled gates, conditioned on the label values $\ell$ running from $0$ to $N/d-1$. The target unitary in the block for a given $\ell$ is $\exp(-i \gamma_{p+1} \hat{\mathcal{H}}_\ell[\psi_p])$, with
\begin{align}
\hat{\mathcal{H}}_\ell[\psi_p] = & \sum_{\substack{i<j\\ \ell_i = \ell_j = \ell}} w_{ij} \frac{ \hat Z_{d_i}\hat Z_{d_j} }{\braket{\psi_p \vert \hat P_\ell \vert \psi_p}}
\nonumber\\
+ & \sum_{\substack{i<j\\ \ell_i\neq \ell_j = \ell}} w_{ij} \bar{z}_i \frac{ \hat Z_{d_j} }{\braket{\psi_p \vert \hat P_{\ell_j} \vert \psi_p}} .
\end{align}
This multicontrolled unitary is then decomposed into multicontrolled Z and ZZ gates, as shown in detail in Fig.~\ref{fig:ansatz-implementation}(b). Further, Fig.~\ref{fig:controlled unitary} shows circuit identities to implement multicontrolled unitaries using simpler gates. Ultimately, the recursive reduction in Fig.~\ref{fig:controlled unitary}(b) can be used to express $\exp(-i\gamma_{p+1} \hat{\mathcal{H}}[\psi_p])$ using only CCPHASE gates, which can be compiled to device-native gates using standard circuit identities.

\subsection{Detailed example with $N=4,d=2$}
\label{subsec: circuit implementation special case}
\begin{figure}[!t]
\includegraphics[width=0.8\columnwidth]{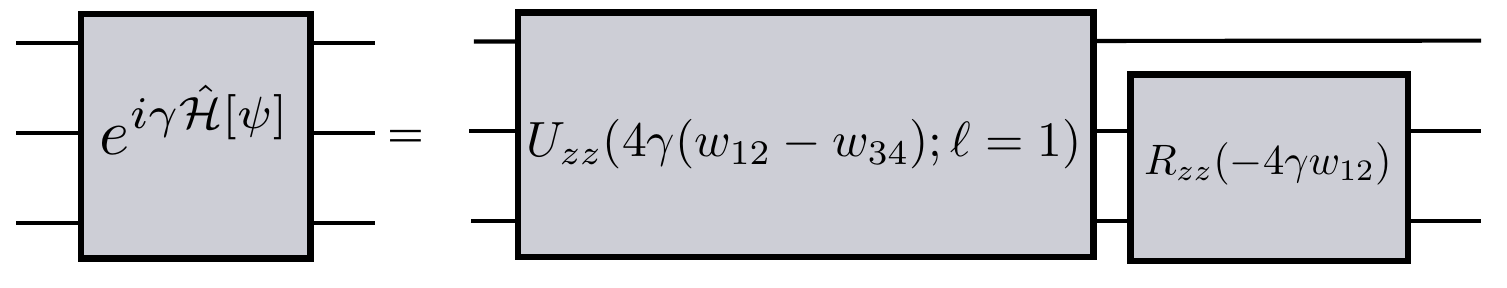}
\caption{Simplification of the phase separator term in the ansatz. Implementation of these logical gates in terms of hardware-native gates are given in Fig.~\ref{circuit_fig_2}}
\label{circuit_fig_1}
\end{figure}
Here, we describe how to implement the phase separator $e^{-i\gamma\hat{\mathcal{H}}[\psi_{p-1}]}$ for the specific case of $N=4,d=2$. In the first layer, we have $\bar{z}_i = 0$ and $\braket{\psi_{p=0} \vert \hat P_\ell \vert \psi_{p=0}} = \frac{1}{2}$. This means that the phase separator Hamiltonian is [see Eq.~\eqref{eq:H special case}]
\begin{equation}
\hat{\mathcal{H}} = 2w_{12} \hat P_{\ell=0} \hat Z_{d_1}\hat Z_{d_2} + 2w_{34} \hat P_{\ell=1} \hat Z_{d_3}\hat Z_{d_4}.
\end{equation}

Next, we assign these qubits to physical qubits on the hardware. Let the label qubit be indexed ``0''. There are two data qubits. Let the physical qubit that encodes the data in variables $z_1$ and $z_3$ be indexed ``1'', i.e. $d_1 = d_3 = 1$. Similarly, let the physical qubit that encodes the data in variables $z_2$ and $z_4$ be indexed ``2'', i.e. $d_2 = d_4 = 2$. The projection operators $\hat P_\ell$ are, respectively,
\begin{align}
\hat P_{\ell=0} = \frac{1 + Z_0}{2}, \nonumber\\
\hat P_{\ell=1} = \frac{1 - Z_0}{2}.
\end{align}
Then, $\hat{\mathcal{H}}$ simplifies to
\begin{align} \label{eq:H special case breakdown}
\hat{\mathcal{H}} = & 2w_{12} \frac{1 + Z_0}{2} \hat Z_1\hat Z_2 + 2w_{34} \frac{1 - Z_0}{2} \hat Z_1\hat Z_2 \nonumber\\
= & 2(w_{34} - w_{12}) \frac{1 - Z_0}{2} \hat Z_1\hat Z_2 + 2w_{12} Z_1 Z_2.
\end{align}
The circuit implementation of the unitary generated by this Hamiltonian is shown in Fig.~\ref{circuit_fig_1}. The unitary generated by the first term of Eq.~\eqref{eq:H special case breakdown} is a controlled-ZZ rotation, and the unitary generated by the second term is a $R_{zz}$ rotation.

\begin{figure}[!t]
\includegraphics[width=1.0\columnwidth]{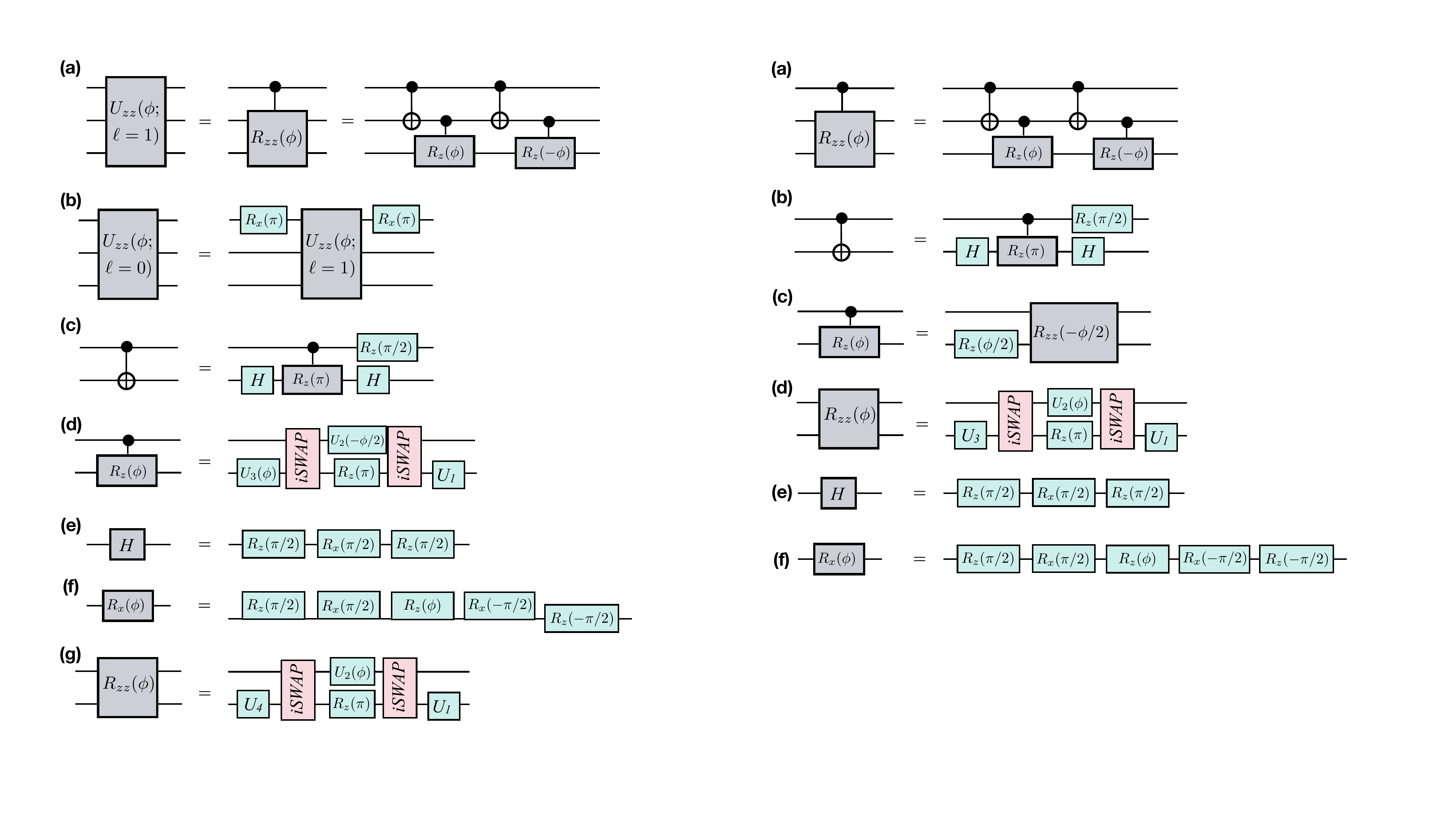}
\caption{Compilation of common gates in terms of gates native to Rigetti's Ankaa\textsuperscript{TM}-9Q-3.}
\label{circuit_fig_2}
\end{figure}

Figure~\ref{circuit_fig_2} shows how to compile $R_{zz}$ and controlled-$R_{zz}$ gates into the native gates on Rigetti's Ankaa\textsuperscript{TM}-9Q-3 chip, $R_x(\pi/2) \equiv \exp(-i\hat X\pi/4)$ and $\textrm{iSWAP} = \exp(i\pi/4(\hat X\otimes\hat X+\hat Y\otimes\hat Y))$. $R_z(\phi)$ gates are virtually applied via frame tracking. The unitary $e^{-i\beta\hat{\mathcal{H}}_x}$ can be implemented with $R_x(2\beta)$ on all the qubits. In Fig.~\ref{circuit_fig_2},
\begin{align}
&U_1 = R_x(-\pi/2)\nonumber\\
&U_2(\phi) = R_z(\pi/2)R_x(\pi/2) R_z(-\phi) R_x(-\pi/2)\nonumber\\
&U_3 = R_z(\pi/2)R_x(\pi/2)
\end{align}

\section{Baseline performance of a related ansatz}\label{sec: baseline}

\begin{figure}[!t]
\includegraphics[width=0.8\columnwidth]{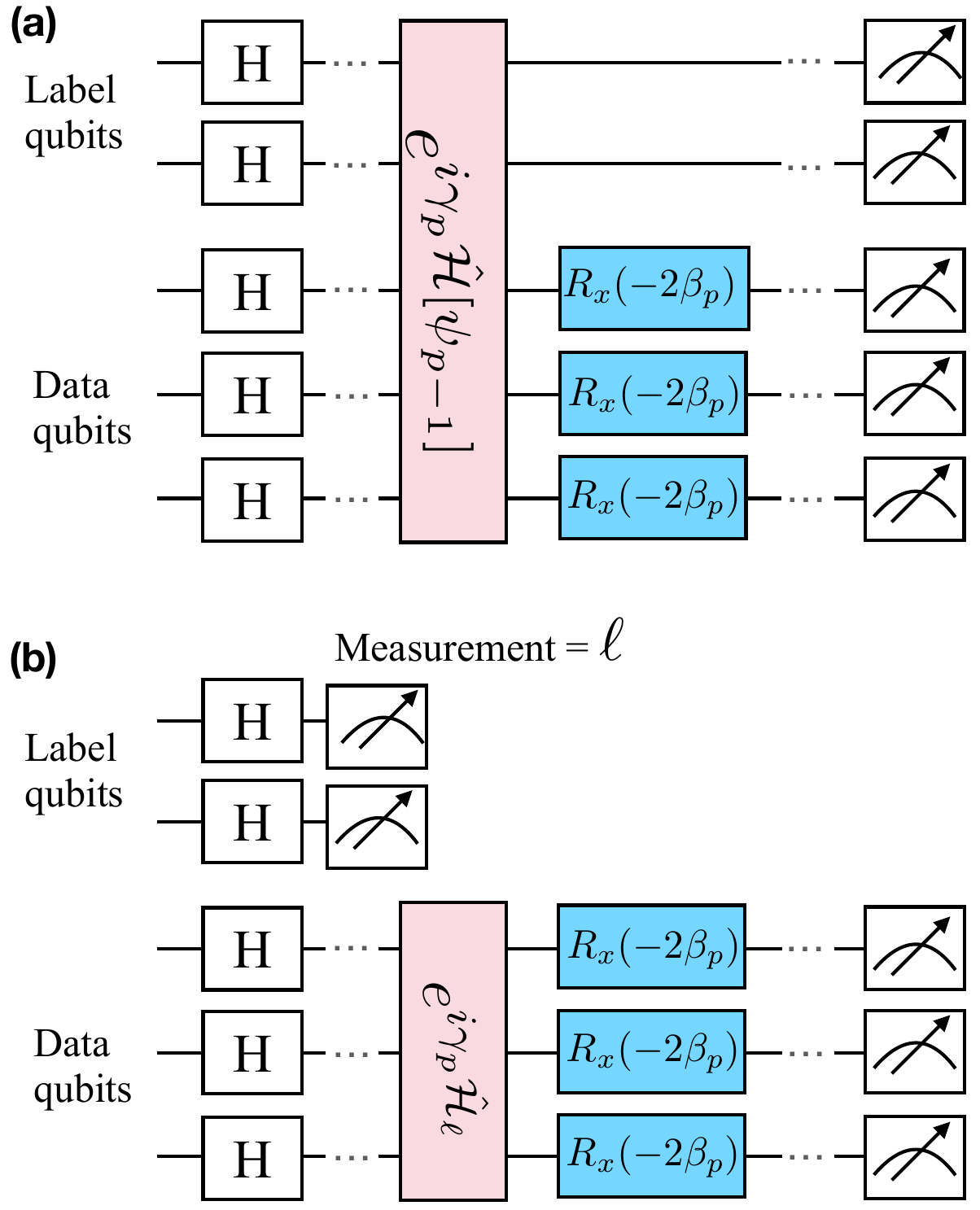}
\caption{An ansatz with phase separators and one-qubit rotations on data qubits only. Executing the ansatz in (a) is equivalent to repeating the ansatz in (b) for $\ell=0\cdots (N/d-1)$, and reduces to implementing QAOA on the clusters with $d$ variables each.}
\label{fig:baseline}
\end{figure}

Here, we consider a slightly modified ansatz given in Fig.~\ref{fig:baseline}(a) The difference with the ansatz in the main text is that there are no $R_x$ rotations of the label bits. Figure~\ref{fig:baseline}(a) can be simplified to the ansatz in Fig.~\ref{fig:baseline}(b). This simplification uses the identity that applying a controlled gate -- apply $U$ on target bits if control bits' value is $x$ -- followed by a measurement of the control bits is equivalent to first doing a measurement on the control bits and then implementing $U$ on the target bits if the measurement is $x$. The ansatz in Fig.~\ref{fig:baseline}(b) essentially implements the QAOA ansatz on $d$ data bits using the SK problem weights on the subset of the $d$ variables corresponding to the measured label value. Repeating the ansatz in Fig.~\ref{fig:baseline}(b) several times essentially solves all the $d$-variable subproblems with label $\ell$, obtains the cost $C_\ell$ for the subproblems $\ell = 0\cdots (N/d-1)$, and returns the final cost,
\begin{equation}
C = \sum_{\ell=0}^{N/d-1} C_\ell.
\end{equation}

When the size of the problem being solved by QAOA is large, Ref.~\cite{farhi2022quantum} predicts clustering of QAOA parameters and tabulates the approximation ratio, $r^*(p)$ up to $p=11$. The asymptotic value of the ground-state cost for $N$ variables is $C^*_{\textrm{asymp.}}(N) = -P N^{3/2} + \alpha N^\omega$, where $P\approx0.7632$ is the Parisi constant~\cite{parisi1979infinite}, $\omega=5/6$ and $\alpha\approx0.7$ is a nonuniversal constant that accounts for finite-size effects~\cite{boettcher2005extremal}. Similarly, for $d\gg1$, the ground-state cost for the subproblems with $d$ variables is $C^*_{\textrm{asymp.}}(d) \sim -P d^{3/2}$, neglecting the subleading term, and the ansatz in Fig.~\ref{fig:baseline} at $p$ layers yields
$C_\ell^*(p) = r^*(p) \times Pd^{3/2}$. Then, the total cost for $N$ variables is
\begin{equation}
C^* = r^*(p)\times P d^{3/2}\times \frac{N}{d}.
\end{equation}
The approximation ratio for this ansatz is
\begin{equation}
r^* \equiv C^*/C^*_{\textrm{asymp.}}(N) \sim r^*(p)\times \sqrt{\frac{d}{N}}.
\end{equation}
This approximation ratio is plotted as the black dash-dotted lines in Fig.~\ref{fig:concentration}.

\section{Shot noise}
\begin{figure}[!t]
\includegraphics[width=1.0\columnwidth]{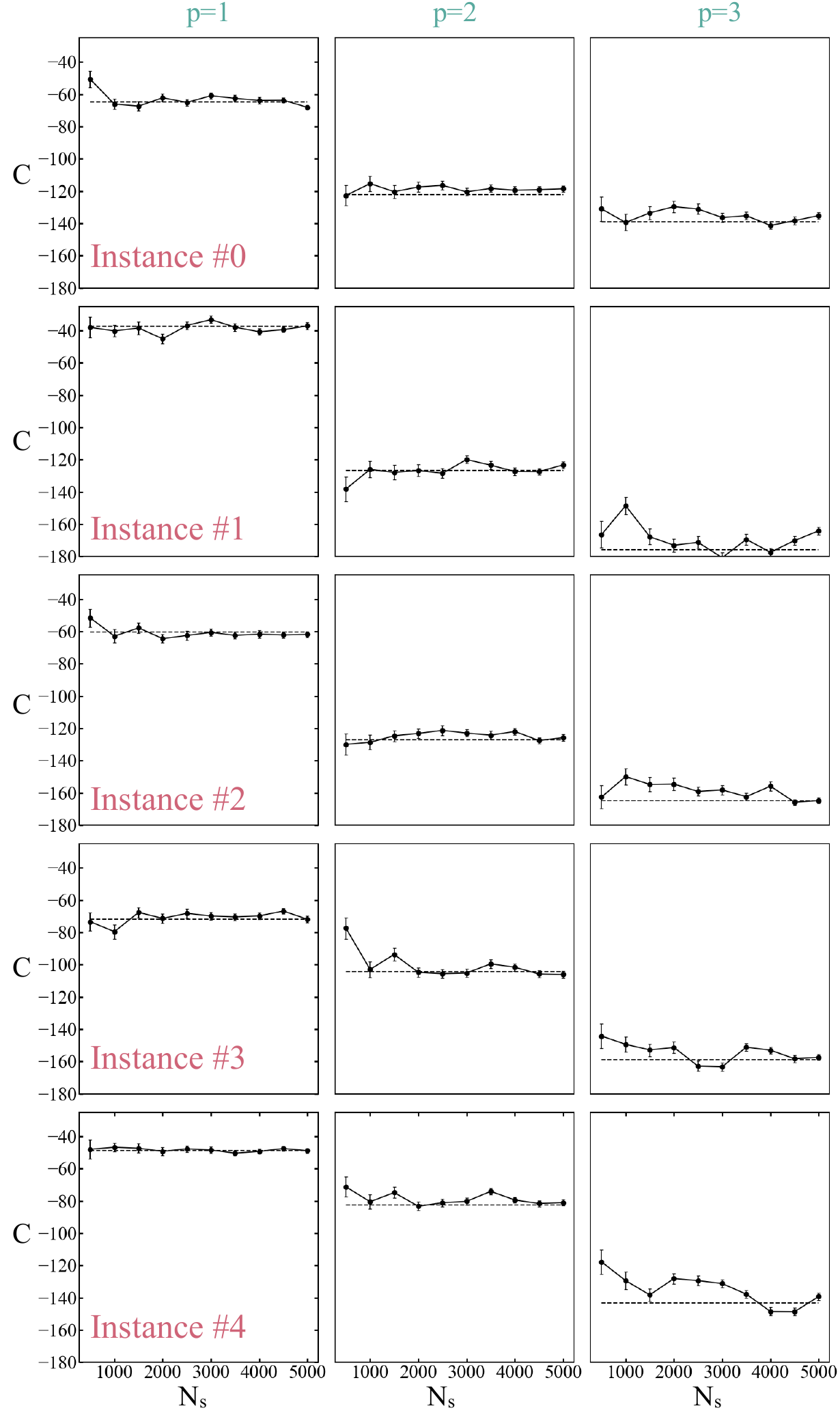}
\caption{Estimated cost versus number of shots for the five SK problem instances in Fig.~\ref{fig:result} at $N=64, d=4$. The costs converge to the exact cost computed from the wave function (dashed line) at $\sim1000$ shots.}
\label{fig:shot_noise_64}
\end{figure}

\begin{figure}[!t]\centering
\includegraphics[width=0.8\columnwidth]{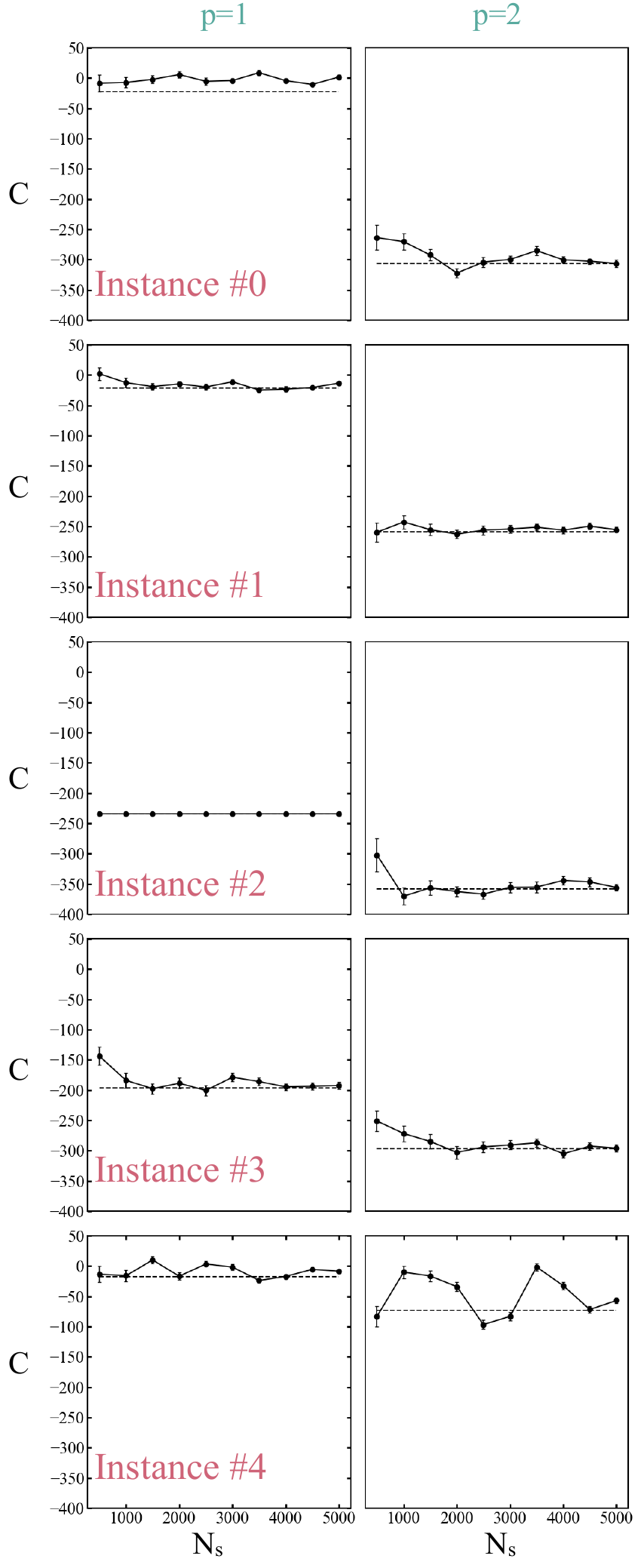}
\caption{Estimated cost versus number of shots for the five SK problem instances in Fig.~\ref{fig:result} at $N=128, d=4$. The costs converge to the exact cost computed from the wave function (dashed line) at $O(1000)$ shots.}
\label{fig:shot_noise_128}
\end{figure}

In Fig.~\ref{fig:result}, we variationally optimized the cost computed from a noiseless calculation of the wave function using a state vector simulator. In Figs.~\ref{fig:shot_noise_64} and~\ref{fig:shot_noise_128}, we show that the estimated cost converges to the exact cost from the wave function at $O(1000)$ shots for $N=64$ and $128$.


\bibliography{bibliography}

\end{document}